\documentclass[11pt,a4paper]{article}
\pdfoutput=1

\usepackage{jcappub}

\usepackage{mathrsfs}
\usepackage{bbm}
\usepackage{bm}
\usepackage{dsfont}
\usepackage{tensor}
\usepackage{xspace}
\usepackage{aeguill}
\usepackage{wasysym}
\usepackage{microtype}
\usepackage{empheq}
\usepackage{ifthen}
\usepackage{amsmath}
\usepackage{subcaption}
\usepackage{wrapfig}
\newcommand{\cst}{\mathrm{cst}}

\newcommand\define{\equiv}

\newcommand\vect[1]{\boldsymbol{#1}}

\newcommand\ex[1]{\mathrm{e}^{#1}}
\renewcommand\i{\ensuremath{\mathrm{i}}}

\newcommand\e[1]{_{\text{#1}}}
\newcommand\h[1]{^{\text{#1}}}
\newcommand\U[1]{\:\mathrm{#1}}

\newcommand{\dd}{\mathrm{d}}

\newcommand{\ddf}[3][]{\frac{\dd^{#1} #2}{\dd {#3}^{#1}}}

\renewcommand\lim[2]{\underset{ #1 \rightarrow #2 }{ \mathrm{lim} } \,}

\newcommand{\delimiters}[4][]{
\ifthenelse{ \equal{#1}{1} }{  #2 #3 #4  }
					{ \ifthenelse{\equal{#1}{2}}{ \big#2 #3 \big#4 }
						{ \ifthenelse{\equal{#1}{3}}{ \Big#2 #3 \Big#4 }
							{ \ifthenelse{\equal{#1}{4}}{ \bigg#2 #3 \bigg#4 }
								{ \ifthenelse{\equal{#1}{5}}{ \Bigg#2 #3 \Bigg#4 }
									{ \left#2 #3 \right#4 }
								}
							}
						}
					}
													}

\newcommand{\pa}[2][]{\delimiters[#1]{(}{#2}{)}}
\newcommand{\pac}[2][]{\delimiters[#1]{[}{#2}{]}}

\newcommand{\ev}[2][]{\delimiters[#1]{\langle}{#2}{\rangle}}

\newcommand{\Hc}{\mathcal{H}}

\newcommand\ee{\end{equation}}
\newcommand\be{\begin{equation}}
\newcommand\eea{\end{eqnarray}}
\newcommand\bea{\begin{eqnarray}}
\newcommand{\bV}{\mathbf{V}}

\newcommand{\bn}{\mathbf{n}}
\newcommand{\bx}{\mathbf{x}}
\newcommand{\by}{\mathbf{y}}
\newcommand{\bk}{\mathbf{k}}  
\newcommand{\te}{\Theta}
\newcommand{\ga}{\Gamma}
\newcommand{\B}{\textrm{B}}
\newcommand{\F}{\textrm{F}}
\newcommand{\X}{\textrm{X}}
\newcommand{\mph}{\,{\rm Mpc}/h}

\newlength{\boxtitlelength}
\newlength{\halfrulelength}
\newcommand{\boxtitle}[1]{\footnotesize\bf{\:#1\:}}

\definecolor{blue4}{RGB}{0,0,143}
\definecolor{red4}{RGB}{143,0,0}
\definecolor{orange}{RGB}{255,128,0}
\definecolor{darkcyan}{RGB}{0,128,128}
\definecolor{olive}{RGB}{0,128,0}
\definecolor{purple}{RGB}{128,0,128}
\definecolor{cyan2}{RGB}{0,255,255}
\definecolor{fushia}{RGB}{255,0,255}
\definecolor{mygray}{gray}{0.5}
\definecolor{lightgray}{gray}{0.85}

\renewcommand{\bk}{\vect{k}}
\renewcommand{\bn}{\vect{n}}
\renewcommand{\bV}{\vect{V}}
\renewcommand{\bx}{\vect{x}}

\makeatletter
\def\@fpheader{\relax}
\makeatother

\title{Testing the equivalence principle on cosmological scales}

\author{Camille Bonvin,}
\author{Pierre Fleury}

\affiliation{D\'{e}partment de Physique Th\'{e}orique, Universit\'{e} de Gen\`{e}ve,\\
24 quai Ernest-Ansermet, 1211 Gen\`{e}ve 4, Switzerland}

\emailAdd{camille.bonvin@unige.ch}
\emailAdd{pierre.fleury@unige.ch}

\abstract{The equivalence principle, that is one of the main pillars of general relativity, is very well tested in the Solar system; however, its validity is more uncertain on cosmological scales, or when dark matter is concerned. This article shows that relativistic effects in the large-scale structure can be used to directly test whether dark matter satisfies Euler's equation, i.e.\,whether its free fall is characterised by geodesic motion, just like baryons and light. After having proposed a general parametrisation for deviations from Euler's equation, we perform Fisher-matrix forecasts for future surveys like DESI and the SKA, and show that such deviations can be constrained with a precision of order 10\%. Deviations from Euler's equation cannot be tested directly with standard methods like redshift-space distortions and gravitational lensing, since these observables are not sensitive to the time component of the metric. Our analysis shows therefore that relativistic effects bring new and complementary constraints to alternative theories of gravity.}


\date{\today}

\begin{document}

\maketitle
\flushbottom

%%%%%%%%%%%%%%%%%%%%%%%%%%%%
\section{Introduction}
\label{sec:introduction}
%%%%%%%%%%%%%%%%%%%%%%%%%%%%

Physical cosmology has reached a level of precision that allows us to test fundamental physics at the percent order, at scales of distance and energy far beyond any terrestrial experiment. After the era of the cosmic microwave background surveys, whose pinnacle was reached with \emph{Planck}~\cite{Adam:2015rua}, comes the time of very large galaxy surveys, such as DESI~\cite{Aghamousa:2016zmz}, Euclid~\cite{Laureijs:2011gra}, the Square Kilometer Array (SKA)~\cite{Braun:2015zta}, or the Large Synoptic Survey Telescope~\cite{Abell:2009aa}. Galaxy surveys offer, in particular, an unprecedented insight into gravitational physics, and have the potential to uncover departures from general relativity (GR), if any.

One of the fundamental pillars of general relativity is Einstein's equivalence principle (EEP), which states that all non-gravitational phenomena are locally unaffected by gravity, if they are performed in a freely falling frame. A particular consequence of this principle is that everything falls according to the same laws, including light. The EEP thus represents the bound between gravity and the rest of physics; it is extremely well tested in the Solar System. Besides, one of its components, namely local Lorentz invariance, implies the $CPT$ symmetry of particle physics~\cite{1951PhRv...82..664S}, which is also well tested on Earth~\cite{Liberati:2013xla}. Nevertheless, the validity of the EEP is much harder to test on cosmic scales, or when the unknown dark matter is concerned. Yet, such tests are equally important as terrestrial ones, since any deviation from the EEP would dramatically shake the basis of fundamental physics in general.

It is interesting to notice that departures from the EEP are, in fact, rarely considered in cosmological tests of gravitation. Notable exceptions are~\cite{Kehagias:2013rpa,Creminelli:2013nua}, exploiting consistency relations between two- and three-point correlation functions of the matter distribution to test for differences between the way baryons and dark matter fall. In general, however, cosmological tests of gravitation beyond GR do assume that the EEP holds. In an inhomogeneous Universe characterised by scalar perturbations only, there are schematically four degrees of freedom: the matter density contrast~$\delta$, the peculiar velocity potential~$V$ of the matter flow, and the two Bardeen potentials~$\Psi$ and $\Phi$ (see e.g.~\cite{PeterUzan}). In GR, under reasonable assumptions, $\Psi=\Phi$ is related to $\delta$ by the Poisson equation, $\delta$ to $V$ by the continuity equation, and $V$ to $\Psi$ via Euler's equation, which closes the system. Alternative theories of gravity generically modify these four relationships. 

\begin{figure}[!t]
\centering
\includegraphics[width=0.8\columnwidth]{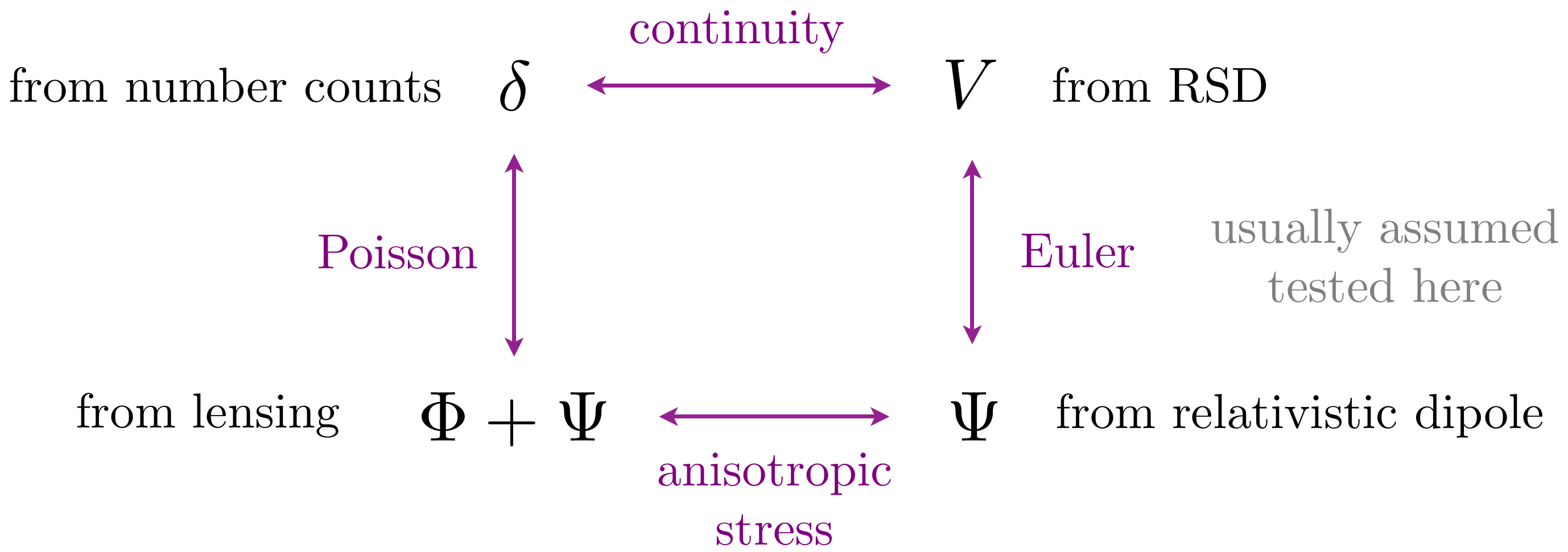}
\caption{Scalar cosmological perturbations consist of four inter-dependent variables: the density contrast~$\delta$, the peculiar velocity potential~$V$, and the gravitational potentials~$\Phi$ and $\Psi$. Standard large-scale structure analyses measure~$\delta$ and $V$ through RSD, and $\Phi+\Psi$ with gravitational lensing. Euler's equation is then used to infer $\Psi$. This allows one to look for deviations from Poisson's equation and for anisotropic stress. These are, therefore, degenerate with violations of Euler's equation. The method proposed in this article is an independent probe of the relationship between $V$ and $\Psi$, which breaks that degeneracy.}
\label{fig:variables}
\end{figure}

The standard way to test for deviations from GR in cosmology consists in combining measurements of redshift-space distortions (RSD) with gravitational lensing (e.g.~\cite{Ferte:2017bpf}). RSD are indeed sensitive to a combination of $\delta$ and $V$, which can be disentangled by measuring both the monopole and quadrupole of the correlation function, whereas gravitational lensing measures the sum of the two metric potentials $\Phi+\Psi$. Therefore, these three measurements allow us to test \emph{three} of the \emph{four} relationships between $\Psi, \Phi, \delta$ and $V$, see fig.~\ref{fig:variables}. 

The most common approach consists in keeping \emph{the continuity and Euler's equations unchanged}\footnote{Here it is implicit that the laws of optics in curved space-time are also left unchanged with respect to GR, so that the whole interpretation of cosmological observations does not need to be modified.} and to test for modifications in the Poisson equation and in the relation between $\Phi$ and $\Psi$~\cite{Amendola:2007rr,Amendola:2012ky}. This has led to precise constraints of the growth rate of structure $f$ (see e.g.~\cite{Alam:2016hwk}) and of the anisotropic stress~\cite{Reyes:2010tr,Ade:2015rim}. 
Relaxing the assumption that Euler's equation holds opens the cosmic Pandora's box, adding a freedom which cannot be constrained by the standard cosmological probes. Indeed, if Euler's equation is allowed to change, then one is left with three probes for four relations. A possible solution is to parameterise modifications to Euler's equation and Einstein's equations with free parameters (usually functions of time) and then use redshift-space distortions and gravitational lensing to constrain these parameters. This method is at the core of the effective field theory of dark energy. However, the fact that we have only three observational probes automatically introduces strong degeneracies between the parameters. This degeneracy cannot be broken without adding extra information about the underlying theory of gravity (like for example assuming a specific form for the Lagrangian). 

In this article, we show that the so-called \emph{relativistic effects} in galaxy surveys are the ideal laboratory to test the EEP \emph{directly}. As such, they can break the degeneracy of the standard set of cosmological probes, induced by potential violations of Euler's equation. Specifically, as originally shown by~\cite{Bonvin:2013ogt}, some relativistic corrections to the observed galaxy number counts generate a dipole in the two-point correlation function. The amplitude of this dipole is affected by \emph{gravitational redshift}, i.e.\,by the gradient of $\Psi$ (the time component of the metric). Combining the dipole with RSD allows us therefore to test directly the relationship between $V$ and $\Psi$ as shown in fig.~\ref{fig:variables}. This would provide the first direct test of the EEP on cosmological scales, since it is sensitive to differences in the way photons and dark matter move\footnote{As explained in more detail in sec.~\ref{sec:theory}, we assume here that galaxies follow the motion of dark matter halos, i.e.\,that there is no velocity bias.}. More precisely, we will see that the relativistic dipole contains terms of the form $(\text{galaxy acceleration} + \nabla \Psi)$, which exactly cancel if the EEP is satisfied, and do not in general. This shows that even though relativistic effects are not significant enough to improve the constraints on parameters that can be measured via standard RSD and lensing techniques (as pointed out in~\cite{Lorenz:2017iez}), they are essential to test for deviations from Euler's equation that would remain unconstrained without them. 

The remainder of the article is organised as follows. In sec.~\ref{sec:theory}, we discuss the notion of free fall in GR and alternative theories of gravity. From the analysis of scalar-tensor and vector-tensor theories, we deduce a quite general parametrisation of the deviations from Euler's equation in cosmology, at sub-Hubble scales. Then, in sec.~\ref{sec:dipole}, we present the relativistic effects in galaxy surveys which give rise to a dipole in the correlation function. We discuss the origin of this dipole, and emphasise its usefulness to constrain the EEP, before explaining how to extract it from the data in practice. In sec.~\ref{sec:forecasts}, we present Fisher-matrix forecasts on deviations from Euler's equation in future surveys like DESI and the SKA. Finally, we conclude in sec.~\ref{sec:conclusion}.

%%%%%%%%%%%%%%%%%%%%%%%%%%%%
\section{Free fall in relativity and beyond}
\label{sec:theory}
%%%%%%%%%%%%%%%%%%%%%%%%%%%%

\subsection{Generalities}
\label{sec:generalities_free_fall}

The curious fact that all things seem to fall the same way has been fundamental in the genesis of Einstein's theory of general relativity (GR)~\cite{Pais}. During the last century, local tests of the weak equivalence principle have reached an exquisite precision~\cite{Will:2014kxa}; lately, the first results of the MICROSCOPE mission~\cite{microscope} have constrained E\"otv\"os ratio at the level of $10^{-14}$~\cite{Touboul:2017grn}. Nevertheless, for obvious reasons, the validity of the equivalence principle is more uncertain on astrophysical and cosmological scales, or when the unknown dark matter is concerned.

In cosmology, if we only consider scalar perturbations of a spatially Euclidean Friedmann-Lema\^itre-Robertson-Walker (FLRW) model in the Poisson gauge, the metric reads
\begin{equation}\label{eq:metric}
\dd s^2 = a^2(\eta) \pac{ -(1+2\Psi)\dd \eta^2 + (1-2\Phi) \delta_{ij} \dd x^i \dd x^j },
\end{equation}
where $a(\eta)$ is the scale factor of the background FLRW model, $\eta$ and $x^i$ being respectively conformal time and spatial comoving coordinates, while $\Psi$ and $\Phi$ are the gauge-invariant Bardeen potentials. If matter is assumed to be freely falling within that space-time, then its flow is characterised by Euler's equation,
\begin{equation}\label{eq:Euler}
\vect{V}' + \Hc \vect{V} + \vect{\nabla}\Psi = 0,
\end{equation}
where $\vect{V}$ is the peculiar matter velocity field, such that the matter four-velocity reads~$v^\mu=a^{-1}(1-\Psi, V^i)$, $\Hc\define a'/a$ is the conformal Hubble rate, and a prime denotes a derivative with respect to conformal time~$\eta$. Equation~\eqref{eq:Euler} directly follows from the geodesic equation, assuming that $|\vect{V}|, \Psi\ll 1$. There are, of course, standard corrections to eq.~\eqref{eq:Euler}, due to the fact that fluid elements are not exactly in free fall. The most common deviation comes from the velocity dispersion of the fluid, which gives rise to a pressure term; other forms of stress can yield different corrections, like shear viscosity, turning Euler's equation into the Navier-Stokes equation. These effects have been considered in a cosmological context, e.g. in~\cite{Blas:2015tla}, and typically manifest on scales below~$1\U{Mpc}$. In this article, we will consider much larger scales and neglect such departures from free fall.

It is also worth mentioning that, even in GR, free fall does not necessarily implies geodesic motion, on which eq.~\eqref{eq:Euler} is based. Indeed, this implication is valid for test particles, but not exactly for extended self-gravitating bodies, despite the validity of the strong equivalence principle. A first example is the case of spinning objects, whose motion is described by the Mathisson-Papapetrou-Dixon equations~\cite{Mathisson:1937zz, Papapetrou:1951pa,Dixon:1970zza}. In weak fields, the deviation from geodesic motion is encoded in a force arising from the coupling between the angular momentum~$\vect{J}$ of the object and the gravito-magnetic field~$\vect{B}\e{g}$ as
$
\vect{F}\e{MPD} = - \vect{\nabla}(\vect{J}\cdot\vect{B}\e{g})
$~\cite{Chicone:2005jj}.
In cosmology, one has at most $\vect{B}\e{g}\sim 10^{-2} \vect{\nabla}\Psi$~\cite{Adamek:2015eda}, so that the ratio between $\vect{F}\e{MPD}$ and the `Newtonian' force $m\vect{\nabla}\Psi$ is of the order of $10^{-2}k R^2\Omega\sim 10^{-7} (k\times 1\U{Mpc})$, where $k$ is the perturbation mode at hand, while $R, \Omega$ are the size and angular velocity of the galaxy. Another known effect, which can make extended objects deviate from geodesic motion, is the so-called self-force, related to the gravitational radiation emitted by the objects~\cite{Poisson:2003nc}.

The `post-geodesic motion' phenomenology gets infinitely richer as one allows for theories of gravity beyond GR. Adding new degrees of freedom to gravitation, or new interactions to the dark-universe sector, generically yields violations of the equivalence principles, affecting \emph{de facto} the way things fall. Such violations can be classified into three categories:
\begin{enumerate}
\item Violation of the weak equivalence principle. The universality of free fall is automatically broken if gravity contains degrees of freedom that couple differently to various matter species. Different couplings can be either fundamental~\cite{Carroll:2008ub} or the result of screening mechanisms~\cite{Hui:2009kc,Jain:2011ji, Brax:2012gr}. This manifests in the apparition of a \emph{fifth force}, which cannot be absorbed by a simple redefinition of metric (Einstein to Jordan frame). The archetype of this situation is a scalar-tensor theory where the scalar degree of freedom~$\phi$ couples differently to dark matter and to the standard model. We will focus on this in sec.~\ref{sec:scalar-tensor}. Another possibility is the direct coupling between dark and baryonic matter~\cite{PhysRevLett.70.119}.
\item Violation of local Lorentz invariance. It is expected, in particular, that the existence of preferred frames or directions implies that objects with different velocities fall differently. An example often cited in this context is the Einstein-\ae ther theory (see sec.~\ref{sec:Einstein-aether}), which contains an additional vector degree of freedom~$u^\mu$ compared to GR. This field can be thought of as the four-velocity of an \ae ther, defining a preferred frame. 
\item Violation of the strong equivalence principle. The strong equivalence principle is very specific to GR. Apart from Nordstr\o m's theory~\cite{Deruelle:2011wu}, none of the alternative seems to satisfy it. Its violation can manifest as a difference between the inertial mass~$m\e{in}$ and the passive gravitational mass~$m\e{p}$ of a self-gravitating body, depending on its gravitational binding energy~$E\e{g}$. This phenomenon is known as the Nordvedt effect~\cite{PhysRev.169.1014}, and is quantified by the parameter~$\eta\e{N}$, as $m\e{p}/m\e{in}=1-\eta\e{N} E\e{g}/m\e{in}$. For a galaxy, $E\e{g}/m\e{in}\sim 10^{-6}$, so that this effect is very small even if $\eta\e{N}\sim 1$. Tests of the strong equivalence principle using black holes have been proposed recently in~\cite{Hui:2012jb, Sakstein:2017bws}.
\end{enumerate}

We restrict our analysis to models where the particles of the standard model (in particular photons) are minimally coupled to gravity, so that the EEP applies to this matter sector, in agreement with experiments. Only dark matter will be allowed to couple non-minimally to the additional gravitational degrees of freedom. Yet, we will assume that the motion of dark matter is directly traced by the motion of galaxies; in other words, \emph{there is no velocity bias}, $V=V\e{g}$. This assumption could seem to be at odds with the fact that we are precisely looking for deviations from the equivalence principle. However, since a galaxy always sits inside a dark matter halo, the latter exerts a binding force on the former, which is very likely to dominate any difference in the way baryonic and dark matter experience gravitation. Such a difference would simply lead to a shift between the baryonic and dark-matter centres of mass~\cite{PhysRevLett.67.2926}, which has been used in~\cite{Desmond:2018euk} to constrain the existence of a fifth force in the local Universe. As such, the method proposed in the present article to test the EEP is not based on the difference between the motion of dark matter and baryons, contrary to~\cite{Creminelli:2013nua}. It is instead based on the difference between the motion of dark matter and photons.

Based on these considerations, we parameterise modifications from Euler's equation~\eqref{eq:Euler} in the following way
\begin{empheq}[box=\fbox]{equation}\label{eq:parametrisation_modified_Euler}
\bV' + \Hc \pac{1+\te(\eta)} \bV +\pac{1+\ga(\eta)} \vect{\nabla}\Psi=0\, .
\end{empheq}
Here $\Theta$ and $\Gamma$ are two free functions of time that encode modifications in the way dark matter (and consequently galaxies) fall in the gravitational potential $\Psi$. The aim of this paper is to constrain these free functions using relativistic effects. Equation~\eqref{eq:parametrisation_modified_Euler} is quite general and contains a rich phenomenology. The parameter $\Gamma$ encodes the effect of  a fifth force acting on dark matter. The parameter $\Theta$ can be thought of as a friction term, which modifies the way the velocity redshifts away. 
As we will see, the specificity of relativistic effects is that they can constrain $\Theta$ and $\Gamma$ independently of the underlying theory of gravity which generates these modifications. 

Before forecasting the constraints on $\Theta$ and $\Gamma$ that we expect from future surveys, let us briefly present two cases which exist in the literature and which generate precisely the kind of deviations written in eq.~\eqref{eq:parametrisation_modified_Euler}.

\subsection{Scalar-tensor theories}
\label{sec:scalar-tensor}

We first focus on a popular class of alternative theories of gravity/dark energy, namely scalar-tensor theories. We start with a simple example (sec.~\ref{sec:scalar-tensor_simple_example}) in order to get an intuition of how the presence of a scalar degree of freedom affects free fall. Despite its simplicity, this example will turn out to essentially contain the physics of the general case (sec.~\ref{sec:scalar-tensor_general}).

\subsubsection{A simple example}
\label{sec:scalar-tensor_simple_example}

Let us consider the simple case of a scalar field~$\phi$, mediating a fifth force via conformal coupling to dark matter only. The corresponding action is
\begin{equation}
S = S\e{EH}[g_{\mu\nu}] 
	+ S\e{SM}[\text{standard matter},g_{\mu\nu}]
	+ S_\phi[\phi,g_{\mu\nu}]
	+ S\e{DM}[\text{dark matter},C^2(\phi)g_{\mu\nu}],
\end{equation}
where $g_{\mu\nu}$ denotes the space-time metric; $S\e{EH}, S\e{SM}$ are respectively the Einstein-Hilbert action and the action of the standard model of particle physics; $S_\phi$ is the canonical action of a scalar field with a potential $U(\phi)$; and finally $S\e{DM}$ is the action of dark matter, which is coupled to the metric via a conformal factor~$C(\phi)$. We model dark matter as a set of spin-less point particles, with individual action
\begin{equation}
S_1 = -m \int |C(\phi)| \; \dd\tau,
\end{equation}
if $\tau$ denotes proper time with respect to the physical metric~$g_{\mu\nu}$, and $m$ is the bare mass of the dark matter particle.

For this model, the equations of motions for $g_{\mu\nu}$, $\phi$, and dark matter are
\begin{align}
R_{\mu\nu} -\frac{1}{2} R g_{\mu\nu} &= 8\pi G \pa{T_{\mu\nu}\h{SM} + T_{\mu\nu}^\phi + |C(\phi)| T_{\mu\nu}\h{DM}} 
\label{eq:EFE_simple_example}\\
\nabla^\mu \nabla_\mu \phi - U_{,\phi} &= C_{,\phi} \, T\e{DM} \label{eq:KG_simple_example} \\
\nabla_\mu(\rho\e{DM} v^\mu) &= 0 \label{eq:conservation_energy_DM}\\
v^\nu \nabla_\nu \pac{ C(\phi) v^\mu} &= - \partial^\mu C \label{eq:fifth_force_simple_example}.
\end{align}
Let us comment on the notation. We chose to call $T^{\mu\nu}\e{DM}=\rho\e{DM}v^\mu v^\nu$ the stress-energy tensor of (cold) dark matter \emph{in the absence of conformal coupling}, and $T\e{DM}\define g_{\mu\nu} T^{\mu\nu}\e{DM}$; $\rho\e{DM}$ must be thought of as related to the number of dark matter particles, while $\vect{v}$ is the four-velocity of the dark matter flow. Note that $\rho\e{DM}$ is conserved, by virtue of eq.~\eqref{eq:conservation_energy_DM}.

The presence of the conformal coupling between dark matter and $\phi$ has three physical effects:
\begin{enumerate}
\item It changes the dark matter \emph{active gravitational mass} by a factor $|C(\phi)|$, as this factor multiplies the bare stress-energy tensor~$T_{\mu\nu}\h{DM}$ in eq.~\eqref{eq:EFE_simple_example}.
\item It also changes its \emph{passive gravitational mass}  and \emph{inertial mass} by a factor $C(\phi)$ ---which is assumed to be positive here--- as seen in the left-hand side of eq.~\eqref{eq:fifth_force_simple_example}.
\item It adds a fifth force proportional to the gradient $\partial^\mu\phi$ of the scalar field%
\footnote{
\label{footnote:friction}
Expanding the left-hand side of eq.~\eqref{eq:fifth_force_simple_example}, one can rewrite it as
\begin{equation*}
v^\nu \nabla_\nu v^\mu = -\partial^\mu_\perp \ln C,
\end{equation*}
where $\partial^\mu_\perp = (\delta^\mu_\nu + v^\mu v_\nu) \partial^\nu$ is the spatial gradient of $\phi$ in the dark matter frame. This alternative expression has the advantage to show that the effect of the fifth force is frame dependent. Suppose that there exists a frame in which $\phi$ is homogeneous (but not static). If a dark matter particle is at rest with respect to this homogeneity frame, then it experiences no fifth force, since $\partial^\mu_\perp\phi=0$. However, if the particle has a small velocity~$v^i$ with respect to that frame, then the spatial gradient becomes $\partial^i_\perp \phi \approx -\dot{\phi} v$, that is a friction force.}.
This gradient, in turn, is sourced by dark matter via $T\e{DM}$ in eq.~\eqref{eq:KG_simple_example}. Note that, contrary to gravitation, the fifth force is \emph{weaker} if it is generated by a hotter dark matter fluid.
\end{enumerate}

In cosmology, if we write~$\phi=\bar{\phi}(\eta)+\delta\phi$,  where $\bar{\phi}$ is the background value of the scalar field, the dark matter equation of motion~\eqref{eq:fifth_force_simple_example} becomes
\begin{equation}\label{eq:modified_Euler_simple_example}
V' + \Hc \pac{ 1+ \frac{C_{,\phi}}{C}\frac{\bar{\phi}'}{\Hc} } V + \Psi = - \frac{C_{,\phi}}{C} \; \delta\phi .
\end{equation}
One can then use the other field equations to establish a relationship between  $\delta\phi$ and $\Psi$. In the quasi-static approximation, and assuming that dark matter dominates as a source of gravitation, one finds~$\delta\phi=(C_{,\phi}/4\pi G) \Psi$. The modified Euler equation then indeed takes the same form as eq.~\eqref{eq:parametrisation_modified_Euler},
\begin{equation}
V' + \Hc \pac{ 1+ \frac{C_{,\phi}}{C}\frac{\bar{\phi}'}{\Hc} } V + \pac{ 1 + \frac{C_{,\phi}^2}{4\pi G C} }\Psi = 0.
\end{equation}

\subsubsection{General case: Horndeski theories}
\label{sec:scalar-tensor_general}

In the most general formulation of scalar-tensor theories, the scalar field~$\phi$ can also be non-minimally coupled to space-time geometry, generating the broad class of Horndeski theories~\cite{Horndeski1974,Deffayet:2009wt} and beyond~\cite{Gleyzes:2014dya,Gleyzes:2014qga,Zumalacarregui:2013pma}. Besides, every matter species can be in principle conformally and disformally coupled to gravitation. The cosmological behaviour of such theories is conveniently described within the effective-field theory (EFT) approach~\cite{Gleyzes:2014rba,Gleyzes:2015pma}, where the coupling between $\phi$ and gravity is characterised by four (Horndeski) or five (beyond Horndeski) functions of time\footnote{The large freedom a priori allowed within this class of models has been significantly reduced with the recent detection of gravitational waves with an optical countrepart~\cite{TheLIGOScientific:2017qsa,Goldstein:2017mmi,Savchenko:2017ffs}, with strong implications for cosmology~\cite{Lombriser:2015sxa,Lombriser:2016yzn,Ezquiaga:2017ekz, Creminelli:2017sry, Sakstein:2017xjx, Baker:2017hug, Langlois:2017dyl}}, while the coupling between $\phi$ and any matter species is given by two additional functions. Here we follow~\cite{Gleyzes:2015pma}, assuming that baryonic matter is universally coupled to gravity, so that one can choose to work in the associated Jordan frame. Only dark matter is then directly coupled to $\phi$, conformally and disformally.

In this context, the modified Euler's equation is found to be~\cite{Gleyzes:2015pma}\footnote{Our notation for $\Phi, \Psi$ is inverted compared to~\cite{Gleyzes:2015pma}, see eq.~\eqref{eq:metric}.}
\begin{equation}\label{eq:Euler_EFT}
V'+ \Hc(1+3\gamma\e{c}\bar{\phi}') V + \Psi =  -3 \gamma\e{c} \Hc \delta\phi,
\end{equation}
which is identical to eq.~\eqref{eq:modified_Euler_simple_example}, except that $C_{,\phi}/C(\phi)$ is now replaced by $3\Hc \gamma\e{c}$, where $\gamma\e{c}(\eta)$ fully encodes the coupling between $\phi$ and dark matter. Reducing the analysis to the class of Horndeski theories, one can then perform a similar operation as in sec.~\ref{sec:scalar-tensor_simple_example}, and express~$\delta\phi$ as a function of $\Psi$. In the quasi-static limit, and assuming that the energy density of dark matter completely dominates the energy density of baryons, we find
\begin{equation}
\ga \define
\frac{3 \Hc \gamma\e{c}\delta\phi}{\Psi}
= \frac{\beta_\gamma(\beta_\xi+\beta_\gamma)}{1 +\beta_\xi(\beta_\xi+\beta_\gamma)},
\end{equation}
where $\beta_\gamma\propto \gamma\e{c}$, while $\beta_\xi$ is related to the coupling functions of the gravitational sector; see sec.~4 of~\cite{Gleyzes:2015pma} for further details. Calling~$\te\define 3\gamma\e{c}\bar{\phi}'$, we recover the phenomenological form~\eqref{eq:parametrisation_modified_Euler} of the modified Euler's equation.

\subsection{Vector-tensor theories}
\label{sec:Einstein-aether}

As a further step towards generality, this section deals with a class of vector-tensor theories, namely Einstein-\ae ther theories, with direct coupling between dark matter and \ae ther. We are going to show that, under some reasonable assumptions, the modified Euler equation takes the same form as in eq.~\eqref{eq:parametrisation_modified_Euler}.

\subsubsection{Fundamentals}

The most natural way to break Lorentz invariance in gravitation, while keeping general covariance, consists in exhuming the idea of an \ae ther, defining a preferred frame. This idea is implemented by equipping gravity with an extra vector degree of freedom~$u^\mu$, which must be thought of as the four-velocity of \ae ther. Following refs.~\cite{Jacobson:2000xp,2008arXiv0801.1547J}, we consider the action
\begin{equation}\label{eq:action_Einstein_aether}
S[g_{\mu\nu},u^\mu] = \frac{1}{16\pi G} \int  \dd^4 x \sqrt{-g} 
						\pac{ R + K\indices{^\mu^\nu_\rho_\sigma} \nabla_\mu u^\rho \nabla_\nu u^\sigma
								+ \lambda (u^\mu u_\mu + 1) }
\end{equation}
which contains the usual Einstein-Hilbert term, but also the kinetic term for $u^\mu$, with
\begin{equation}
K_{\mu\nu\rho\sigma} \define c_1 g_{\mu\nu} g_{\rho\sigma}
										 + c_2 g_{\mu\rho} g_{\nu\sigma} 
										 + c_3 g_{\mu\sigma} g_{\nu\rho}
										 - c_4 u_{\mu} u_{\nu} g_{\rho\sigma}\, ,
\end{equation}
the coefficients~$c_{1\ldots 4}$ being free parameters of the theory\footnote{We use the same convention as refs.~\cite{Will:2014kxa, Audren:2014hza}. Note the difference with refs.~\cite{Jacobson:2000xp,2012JCAP...10..057B}, where mostly-minus signature was used, and $c_4\rightarrow -c_4$. The authors of~\cite{Carroll:2004ai} chose to parameterise the Einstein-\ae ther theory by~$\beta_1, \beta_2, \beta_3$, with $c_a=-16\pi G \beta_a$, and $c_4=0$.}. These parameters are already severely constrained by experiments. On the one hand, tests in the Solar system impose $|\alpha_1|<10^{-4}$ and $|\alpha_2|<2\times 10^{-9}$, where those PPN parameters are given by the combinations~\cite{Will:2014kxa}
\begin{align}
\alpha_1 &\define -\frac{8(c_3^2+c_1c_4)}{2c_1 - c_1^2+c_3^2} \\
\alpha_2 &\define 2\alpha_1 - \frac{[2(c_1+c_3)-(c_1+c_4)][(c_1+c_4)+(c_1 + 3 c_2 + c_3)]}{(c_1+c_2+c_3)(2-c_1-c_4)}.
\end{align}
On the other hand, the recent constraints on the relative velocity of light and gravitational waves set by GW170817~\cite{TheLIGOScientific:2017qsa} and GRB170817A~\cite{TheLIGOScientific:2017qsa,Goldstein:2017mmi,Savchenko:2017ffs} impose $|\alpha\e{T}|<10^{-15}$~\cite{Baker:2017hug}, where
\begin{equation}
\alpha\e{T} \define -\frac{c_1+c_3}{1+c_1+c_3} .
\end{equation}
In the following, we will consider~$\alpha_1=\alpha_2=\alpha\e{T}=0$, which can be satisfied by setting $c_1+c_3=c_1+c_4=0$ if~$c_1\not= 0$; see however~\cite{Oost:2018tcv} for a thorough status of the current constraints on the $c_i$, including constraints from the big-bang nucleosynthesis~\cite{Carroll:2004ai}. The last term in \eqref{eq:action_Einstein_aether} is a constraint which ensures the normalisation~$u_\mu u^\mu = -1$ of the \ae ther four-velocity, while $\lambda$ is a Lagrange multiplier. This theory can be considered a low-energy limit of Ho\v{r}ava gravity~\cite{Horava:2009uw,Blas:2009qj,Blas:2010hb}. Note finally that the action~\eqref{eq:action_Einstein_aether} can be further generalised~\cite{Zlosnik:2006zu,Zhao:2007ce} by replacing the kinetic term by a general function of $K\indices{^\mu^\nu_\rho_\sigma} \nabla_\mu u^\rho \nabla_\nu u^\sigma$.

An explicit violation of the equivalence principle can then be implemented by coupling dark matter particles to $u^\mu$ similarly to how charged particles couple to the electromagnetic four-potential; namely, the action of a dark matter point particle is taken to be~\cite{2012JCAP...10..057B}
\begin{equation}
S_1 = -m \int \dd\tau \; F(\gamma),
\end{equation}
where $\gamma\define -u_\mu v^\mu$ is the relative Lorentz factor between dark matter and \ae ther, and $F$ is a function such that $F(1)=1$. Due to its similarity with the conformal coupling to a scalar field given in sec.~\ref{sec:scalar-tensor_simple_example}, we expect similar phenomena to appear: fifth force, modification of the inertial and gravitational masses, etc.

The full set of equations of motion for $g_{\mu\nu}$, $u^\mu$ and dark matter is rather heavy. We choose to leave it in the appendix~\ref{appendix:Einstein-aether}, while focussing on the dynamics of dark matter for now. If $v^\mu$ denotes the four-velocity of the dark matter flow, and $\rho\e{DM}$ its energy density in the absence of coupling to \ae ther, then
\begin{align}
\nabla_\mu (\rho\e{DM} v^\mu) &= 0 \label{eq:energy_conservation_Einstein-aether}\\
v^\nu \nabla_\nu \pac{(F-\gamma F_{,\gamma})v^\mu}
&= F_{,\gamma} \omega\indices{^\mu_\nu} u^\nu 
	- \dot{\gamma} F_{,\gamma\gamma} v^\mu\, , \label{eq:EOM_DM_Einstein-aether}
\end{align}
where $\dot{\gamma}\define \dd\gamma/\dd\tau$, and $\omega_{\mu\nu}\define \partial_\mu u_\nu - \partial_\nu u_\mu$. As in the scalar-tensor case, eq.~\eqref{eq:energy_conservation_Einstein-aether} tells us that the bare energy of dark matter is conserved. The phenomenology of eq.~\eqref{eq:EOM_DM_Einstein-aether} is richer. On the one hand, the inertial mass is modified by a factor $(F-\gamma F_{,\gamma})$. On the other hand, it experiences two kinds of additional forces. The second term on the right-hand side is a kind of friction, proportional to the relative acceleration between dark matter and \ae ther. The first term is reminiscent of the Lorentz force, $\omega_{\mu\nu}$ being analogous to the field strength of electrodynamics. A more kinetic interpretation consists in viewing $\omega_{\mu\nu}u^\nu$ like an inertial force, containing both dragging-like and Coriolis-like effects. This rich phenomenology will turn out to highly simplify in the context of linear cosmological perturbations.

\subsubsection{Cosmological aspects}

In strictly homogeneous and isotropic cosmology, \ae ther has to be comoving with matter in order to preserve the symmetries of the FLRW space-time. Thus~$\gamma=-u^\mu v_\mu = 1$ and everything goes as if dark matter and \ae ther were uncoupled. The expansion dynamics is nonetheless affected, due to the stress-energy of \ae ther itself. This stress-energy tensor turns out to be directly related to the extrinsic curvature of the homogeneity hypersurfaces, and the net effect is to multiply both $H^2$ and $a^{-1}\dd^2 a/\dd t^2$ in the Friedmann equations\footnote{The effect of \ae ther on the dynamics of cosmic expansion was first considered in~\cite{Carroll:2004ai}, for $c_4=0$ and no cosmological constant. The authors chose to interpret the factor $1-(c_1+3c_2+c_4)/2$ as a renormalisation of Newton's constant and spatial curvature. Had they considered $\Lambda\not=0$, they would have concluded that the cosmological constant had to be renormalised as well.} by~$1-(c_1+3c_2+c_4)/2$.

At the level of perturbations, things are more involved. Restricting to scalar modes as we did in sec.~\ref{sec:scalar-tensor}, the modified Euler equation of dark matter reads
\begin{equation}\label{eq:modified_Euler_Einstein-aether}
V' + \Hc V + \Psi = Y \pac{ V'-U' + \Hc(V-U) },
\end{equation}
where $U$ is the velocity potential associated to $u^i$, and where we used the notation of~\cite{2012JCAP...10..057B}, $Y\define F_{,\gamma}(1)$, for the effective coupling constant between dark matter and \ae ther. The first constraints on $Y$ were performed in~\cite{Audren:2014hza}, who obtained~$Y<3\%$ by combining CMB and BAO data. More recently, a method based on galactic dynamics was proposed in~\cite{2017JCAP...05..024B}. Note that, as $A_V\define V'+\Hc V+\Psi$ represents the 4-acceleration potential of dark matter, eq.~\eqref{eq:modified_Euler_Einstein-aether} can be written as~$A_V=Y(A_V-A_U)$, where we see that the fifth force is related to the relative acceleration of the two flows.

When the above is combined with the dynamics of \ae ther and gravitation (see appendix~\ref{appendix:Einstein-aether}), it can be shown that, in Fourier space, $V-U = h(a,k) V$, with
\begin{equation}
h(a,k) \define \frac{c_2(2-3 c_2) a k^2 + 9 c_2^2 \Omega\e{m0} H_0^2}
								{c_2 (2-3 c_2) a k^2 - \Omega\e{m0} H_0^2 \pac{(2-3 c_2)Y+3c_2(1-c_2/2)}}\, ,
\end{equation}
and where we introduced~$\Omega\e{m}=8\pi G\rho_0/3$, corresponding to the uncoupled dark matter density today, $\rho_0$. We can then substitute in the relative acceleration of the two fluids, $A_V-A_U = Y(V'+\Hc V)+h'V$, and we obtain
\begin{equation}
V' + \Hc \pa{1 - \frac{Y\Hc^{-1}h'}{1-Y h}} V + \pa{ 1 + \frac{Y h}{1-Y h}}\Psi = 0\, ,
\end{equation}
which has the same form as eq.~\eqref{eq:parametrisation_modified_Euler}, except that the functions encoding departures from Euler are now scale dependent (through $h$). However, for sub-Hubble modes~$k \gg \Hc$, $h=1+\mathcal{O}(\Hc^2/k^2)$, and we are left with
\begin{equation}
V' + \Hc V + \pa{ 1+\frac{Y}{1-Y} } \Psi = 0\, .
\end{equation}
up to second order in $\Hc/k$.

\subsection{Discussion}

\label{sec:parametrisation}

The two theoretical cases investigated in secs.~\ref{sec:scalar-tensor} and \ref{sec:Einstein-aether} produce precisely the kind of deviations proposed in eq.~\eqref{eq:parametrisation_modified_Euler}. One could then wonder how general such a parametrisation is. After all, the fact that we only consider linear scalar perturbations should restrict the possible modifications of Euler's equation. An apparently reasonable guess for the latter would be $V'=L(V,\Phi,\Psi,\delta,\psi_i)$, where $\psi_i$ represents the extra degrees of freedom ($\phi$ in the scalar-tensor case, $U$ in the Einstein-\ae ther case, etc.) and where $L$ is linear with respect to its arguments. One would then use the other equations of motion to eliminate~$\Phi, \delta, \psi_i$, so that $V'=A V+ B\Psi$, which strongly resembles~\eqref{eq:parametrisation_modified_Euler}.

One could, however, imagine extensions of this parametrisation. First of all, $L$ could in principle depend on the time derivative of its arguments. Even in the quasi-static approximation, it is not obvious that all those derivatives could be neglected, in particular if they are combined with spatial derivatives. This leads to the second point, which is that the coefficients $A$ and $B$ in $V'=A V+ B\Psi$ could not only be time-dependent, but also generically scale-dependent. In our forecasts we will not consider this possibility, but it could lead to stronger constraints by modifying not only the amplitude of the relativistic correlation function, but also its shape.

Finally, let us note that if $\te\ll\ga$, then $\ga$ essentially coincides with the E\"otv\"os ratio $2(a\e{DM}-a\e{B})/(a\e{DM}+a\e{B})$ between dark and baryonic matter. It is automatically the case in the Einstein-\ae ther model, since~$\te=0$. In the general scalar-tensor scenario described by EFT, one has
\begin{equation}
\frac{\te}{\ga}
= \frac{3\gamma\e{c}\bar{\phi}'}{\Gamma}
\end{equation}
which is small if $\bar{\phi}$ evolves very slowly, but could in principle be of order unity. Physically speaking, it corresponds to the situation where the effect of the running of the dark matter mass is smaller than the effect of the fifth force.

%%%%%%%%%%%%%%%%%%%%%%%%%%%%
\section{Testing Euler's equation with galaxy surveys}
\label{sec:dipole}
%%%%%%%%%%%%%%%%%%%%%%%%%%%%

In the previous section, we motivated the general parametrisation~\eqref{eq:parametrisation_modified_Euler} of deviations of the dark matter flow with respect to Euler's equation. This section deals with the measurability of such deviations. We will now explain how such a signal can be extracted from relativistic effects in galaxy surveys.

\subsection{What galaxy surveys really measure}

\begin{figure}[h!]
\centering
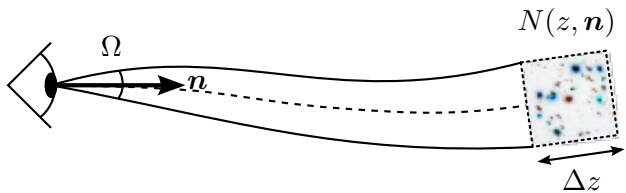
\caption{Galaxy number count~$N(z,\vect{n})$, observed in a pixel~$\Omega$ of the sky about the direction~$\vect{n}$, in a redshift bin~$\Delta z$ about $z$.}
\label{fig:number_counts}
\end{figure}

Galaxy surveys attempt to trace the distribution of matter in the Universe from the number density of galaxies. The main observable is therefore the number~$N$ of galaxies per unit of observed volume, i.e. per pixel of the sky (subtended by a solid angle~$\Omega$) and per redshift bin~$\Delta z$ (see fig.~\ref{fig:number_counts}). The inhomogeneity of the distribution of galaxies is then quantified by
\begin{equation}
\Delta \define \frac{N - \bar{N}}{\bar{N}},
\end{equation}
where $\bar{N}$ is the average of $N$ over all the pixels, i.e. the total number of observed galaxies divided by the volume of the survey.

The observable~$\Delta(z,\vect{n})$ does not only contain information about the matter density contrast around the point $(z,\vect{n})$, but also about the relation between the observed pixel~$(\Delta z, \Omega)$ and the corresponding physical volume. Indeed, the gravitational effects of matter inhomogeneities affect the propagation of light, and its frequency. As a consequence, a given pixel~$(\Delta z, \Omega)$ can be first physically larger or smaller compared to its counterpart in a strictly homogeneous Universe, and second it can be physically closer or further away from the observer than it would be in a homogeneous Universe. These contributions to $\Delta$ have been calculated in~\cite{Yoo:2009au,Bonvin:2011bg, Challinor:2011bk} at linear order in scalar cosmological perturbations. The result is conveniently written as
\begin{equation}
\Delta = \Delta\h{st} + \Delta\h{rel} + \Delta\h{lens}\,,
\end{equation}
where $\Delta\h{st}$ is the standard expression of $\Delta$, used in all galaxy surveys, and which already accounts for the effect of biased tracers and redshift-space distortions; $\Delta\h{rel}$ contains the so-called relativistic effects, which are corrections to the redshift (Doppler and Einstein effects, integrated Sachs-Wolfe effect, \ldots), and hence ensure the correct estimation of the physical depth and width of redshift bins; and finally $\Delta\h{lens}$ denotes the contribution of gravitational lensing, which translates the observed angular size~$\Omega$ of a pixel into physical areas. Their expressions are 
\begin{align}
\Delta\h{st} &= b \delta - \frac{1}{\Hc} \, \partial_r(\vect{V}\cdot\vect{n})\, , \label{eq:Delta_standard}\\
\Delta\h{rel}&=\frac{1}{\Hc}\partial_r\Psi+\left(1-5s+\frac{5s-2}{r\Hc}-\frac{\Hc'}{\Hc^2} \right) \bV\cdot\bn+\frac{1}{\Hc}\bV'\cdot\bn\, ,\label{eq:Delta_relativistic}\\
\Delta\h{lens}&=(5s-2)\int_0^r \dd r' \; \frac{r-r'}{2rr'}\Delta_\Omega(\Phi+\Psi)\, .\label{eq:Delta_lens}
\end{align}
Here $b$ denotes the linear bias of the matter tracer (typically galaxies), and $r$ is the comoving radial coordinate in the direction of~$\vect{n}$; in eq.~\eqref{eq:Delta_relativistic} and~\eqref{eq:Delta_lens}, $s$ denotes the slope of the luminosity function which parameterises the magnification bias and $\Delta_\Omega$ is the transverse Laplacian. Note that, in eq.~\eqref{eq:Delta_relativistic}, we dropped the terms of $\Delta\h{rel}$ involving the gravitational potentials $\Psi$ and $\Phi$ without spatial derivatives, because their contribution to the dipole presented in the next subsection is suppressed by $(\Hc/k)^2$ with respect to the other terms.

Let us further focus on the relativistic correction~$\Delta\h{rel}$ given by eq.~\eqref{eq:Delta_relativistic}. The first term denotes de contribution from gravitational redshift, while the other terms are Doppler effects. The fact that gravitational redshift depends directly on $\Psi$ allows us to test Euler's equation in a model-independent way. We notice that three of its terms exactly cancel if Euler's equation is satisfied. This turns out to be a direct consequence of Einstein's equivalence principle---we refer the interested reader to appendix~\ref{app:equivalence} for more details about this connection. Thus, relativistic effect in galaxy surveys are an ideal laboratory to look for violation of the equivalence principle. If Euler's equation is violated according to our proposition~\eqref{eq:parametrisation_modified_Euler}, then eq.~\eqref{eq:Delta_relativistic} becomes
\begin{equation}
\label{Delta_mod}
\Delta\h{rel}(\bn, z)
=
\pa{\frac{\ga-\te}{1+\ga}-5s+\frac{5s-2}{r\Hc}-\frac{\Hc'}{\Hc^2} } \bV\cdot\bn+\frac{\ga}{\Hc(1+\ga)}\,\bV'\cdot\bn\, ,
\end{equation}
where a couple of terms vanish for $\ga=\te=0$. In the remainder of this section, we show how to extract $\Delta\h{rel}$ from galaxy surveys.

\subsection{Dipolar correlations}
\label{sec:aymm}

\begin{figure}[h!]
\centering
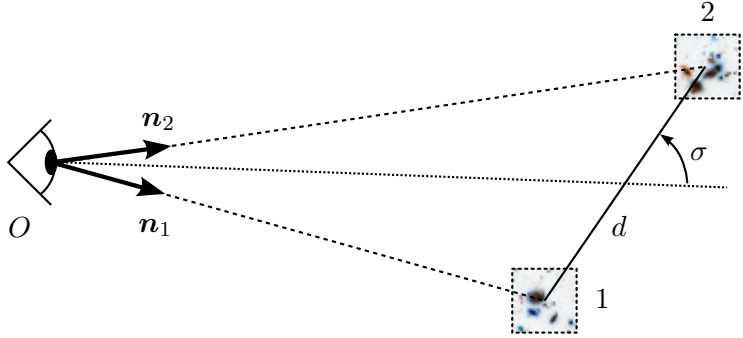
\caption{Relative localisation of two pixels, by their comoving distance~$d$ and the angle~$\sigma$ between the axis $(12)$ and the line of sight $\vect{n}_1$. Note that we used Euclidean geometry on that drawing because the effects of cosmological perturbations are already taken into account in the terms $\Delta\e{rel}$ and $\Delta\e{lens}$ of $\Delta$.}
\label{fig:parametrisation_correlation}
\end{figure}

The simplest way of extracting cosmological information from galaxy surveys consists in using the two-point correlation function of $\Delta$,
\begin{equation}
\xi(z_1,\vect{n}_1;z_2,\vect{n}_2) \define \ev{\Delta(z_1,\vect{n}_1) \Delta(z_2,\vect{n}_2)}.
\end{equation}
Due to statistical homogeneity and isotropy, only three out of the six variables $(z_1,\vect{n}_1,z_2,\vect{n}_2)$ are necessary to parameterise $\xi$, because only the relative position of the two pixels matters. A convenient parametrisation, depicted in fig.~\ref{fig:parametrisation_correlation}, consists in locating, for example, pixel 2 relatively to pixel 1 by their mutual distance~$d$ and the angle~$\sigma$ between the axis $(12)$ and the mean line-of-sight~$\vect{n}$. We thus write~$\xi(d,\sigma,z)$, where $z$ is the redshift of the pixel~$1$. 

A key advantage of this parametrisation is that the various contributions to $\xi(d,\sigma,z)$ depend differently on the angle~$\sigma$. For example, it is well known that redshift-space distortions yield a quadrupole and hexadecapole in terms of~$\sigma$~\cite{Kaiser:1987qv,Hamilton:1997zq}. \emph{It turns out that relativistic effects add a dipole and an octupole to this picture}~\cite{Bonvin:2013ogt,Croft:2013taa}, which makes them identifiable in the data\footnote{Note that in the power spectrum, relativistic effects are identifiable by the fact that they generate an imaginary part~\cite{McDonald:2009ud,Yoo:2012se}.}. Let us briefly summarise why such a dipolar structure appears (see~\cite{Bonvin:2013ogt,Croft:2013taa} for more detail).

As an example, we focus on the correlation between the first term in eq.~\eqref{eq:Delta_standard} (density) and the first term of eq.~\eqref{eq:Delta_relativistic} (gravitational redshift),
\begin{equation}\label{eq:expansion_correlation}
\xi^{b\delta\times\partial_r\Psi}(\vect{x}_1,\vect{x}_2)
= \ev{b\delta(\vect{x}_1)\partial_r\Psi(\vect{x}_2)}+\ev{\partial_r\Psi(\vect{x}_1)b\delta(\vect{x}_2)}\, .
\end{equation}
Each of these terms is antisymmetric with respect to the exchange of $\vect{x}_1,\vect{x}_2$, which means that they have a dipolar component in terms of the angle~$\sigma$ depicted in fig.~\ref{fig:parametrisation_correlation}. Indeed, suppose that there is an over-density at $\vect{x}_1$ [$\delta(\vect{x}_1)>0$]. This generates a potential well around~$\vect{x}_1$, with $\vect{\nabla}\Psi$ directed outwards. Thus, if $\vect{x}_2$ is located behind~$\vect{x}_1$ along the line of sight ($\sigma=0$), then $\delta(\vect{x}_1)\partial_r\Psi(\vect{x}_2)>0$; conversely, if $\vect{x}_2$ is located between $\vect{x}_1$ and the observer ($\sigma=\pi$), then $\delta(\vect{x}_1)\partial_r\Psi(\vect{x}_2)<0$. The same reasoning applies if there is an under-density at $\vect{x}_1$.

\begin{figure}[h!]
\centering
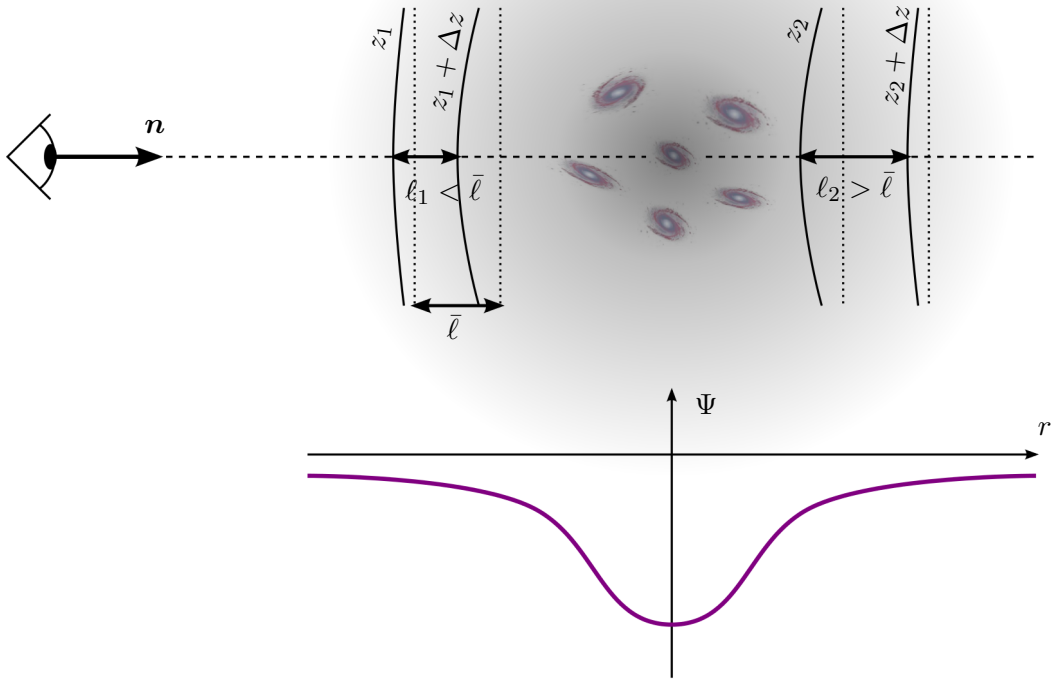
\caption{Effect of the gravitational potential created by an over-density (centre of the black halo) on the physical size of redshift bins located in front of it or behind it. Solid lines correspond to iso-$z$ surfaces, while dotted lines are iso-$r$ surfaces (which would coincide with $z=\cst$ in the FLRW background). A point closer to the over-density experiences a stronger gravitational field, which enhances its redshift; in order to have the same redshift as a point that experiences a weaker gravitational field, it has to be closer to the observer. The net effect is to squeeze (resp. stretch) the redshift bins in front of (resp. behind) the over-density.}
\label{fig:asymmetric}
\end{figure}

This antisymmetry of gravitational redshifts can be understood as in fig.~\ref{fig:asymmetric} (see also fig. 2 of~\cite{Bonvin:2013ogt}). Consider two redshift bins~$\Delta z$ located respectively in front of, and behind, an over-density. The closer a galaxy is to the over-density, the stronger its  gravitational redshift. This distorts the iso-$z$ surfaces in real space, as gravitational redshift mimics the effect of cosmic expansion: the stronger the gravitational field experienced by a galaxy is, the less far from the observer it needs to be in order to have a given redshift. Hence, the physical thickness of the bin in front of the over-density is effectively squeezed, so it potentially contains less galaxies, which reduces~$\Delta$, and conversely for the bin located behind the over-density, whence the dipole. The same kind of reasoning can be made for the effect of velocity and acceleration.

In eq.~\eqref{eq:expansion_correlation}, $\ev{b\delta(\vect{x}_1)\partial_r\Psi(\vect{x}_2)}$ is accompanied with $\ev{\partial_r\Psi(\vect{x}_1)b\delta(\vect{x}_2)}$. It is then not hard to see that the dipoles associated to each term exactly compensate, because exchanging~$\vect{x}_1$ and $\vect{x}_2$ corresponds to changing $\sigma$ into $\sigma+\pi$. However, this cancellation does not happen if one takes the cross-correlation between \emph{two types of tracers, with different biases}. For instance, correlating bright~(B) and faint~(F) galaxies with respective biases~$b\e{B}, b\e{F}$, eq.~\eqref{eq:expansion_correlation} becomes
\begin{equation}
\xi\h{cross}\e{BF}(\vect{x}_1,\vect{x}_2)
= \ev{b\e{B}\delta(\vect{x}_1)\partial_r\Psi(\vect{x}_2)}+\ev{\partial_r\Psi(\vect{x}_1)b\e{F}\delta(\vect{x}_2)}\, .
\end{equation}
%
%
%\begin{align}\label{eq:expansion_correlation_asymmetric}
%\xi\e{BF}(\vect{x}_1,\vect{x}_2)
%&\define \ev{\Delta\e{B}(\vect{x}_1)\Delta\e{F}(\vect{x}_2)} \\
%&= \ev{\Delta\h{st}\e{B}(\vect{x}_1)\Delta\h{st}\e{F}(\vect{x}_2)}
%	\underbrace{
%	+ \ev{\Delta\h{st}\e{B}(\vect{x}_1)\Delta\e{rel}(\vect{x}_2)}
%	+ \ev{\Delta\e{rel}(\vect{x}_1)\Delta\e{F}\h{st}(\vect{x}_2)}
%	}_{\define \xi\e{BF}\h{rel}(\vect{x}_1,\vect{x}_2)}
%	+ \ldots
%\end{align}
%
Both terms still contain opposite dipoles, but the former is proportional to~$b\e{B}$, while the latter is proportional to $b\e{F}$. There is, therefore, a net dipole in $\xi\e{BF}\h{cross}$ proportional to $b\e{B}-b\e{F}$.

\subsection{Extracting the dipole}

Because the standard contributions to the cross-correlation function do not contain any dipolar component in the flat-sky approximation, relativistic effects can be extracted by integrating the latter with a suitable antisymmetric kernel. This subsection summarises the construction of an estimator of the dipole (see~\cite{Bonvin:2015kuc} for more detail).

\subsubsection{Set-up and conventions}

To calculate the multipoles of the correlation function, we write $\Delta$ in Fourier space, for which we use the convention\footnote{$\Delta$ is measured as a function of $z$, but it can now be re-written in terms of conformal time~$\eta$, since the error that is made when one goes from $z$ to $\eta$ using the background relationship has been consistently included in our derivation of $\Delta$. In the following, we will use indistinctly $z$ or $\eta$.}
\begin{equation}
\Delta(\bk, \eta) = \int \dd^3 x \; \ex{\i \bk\cdot\bx}\Delta(\bx, \eta) \, ;
\qquad
\Delta(\bx, \eta) = \int \frac{\dd^3 k}{(2\pi)^3} \; \ex{-\i \bk\cdot\bx}\Delta(\bk, \eta) \, .
\end{equation}
We define the velocity potential $\hat V$ in Fourier space through $\bV(\bk,\eta)\define \i(\vect{k}/k)\hat V(\bk,\eta)$, so that $\hat V(\bk, \eta)$ has the same dimensions as $\delta(\bk, \eta)$, i.e. [length]$^3$. Note that $\hat V(\bk, \eta)$ is then related to the Fourier transform of $V(\bx, \eta)$ by a factor $-k$. We assume that the continuity equation is valid, as discussed in sec.~\ref{sec:theory}. For sub-Hubble modes ($k\gg\Hc$),
\begin{equation}
\label{continuity}
\hat V(\bk, \eta)=-\frac{1}{k}\,\delta'(\bk,\eta)\, .
\end{equation}
This allows us to write $\Delta(\bk, \eta)$ as a function of $\delta(\bk, \eta)$ only, in particular
\begin{align}
\Delta^{\rm st}(\bk, z) &=\pac{ b(z)+f(z)(\hat\bk\cdot\bn)^2 } \delta(\bk, \eta)\ ,\label{Deltast_k}\\
\Delta^{\rm rel}(\bk, z) &=\i\hat\bk\cdot\bn \frac{\Hc}{k}
											\bigg[
													\pa{ 5s+\frac{2-5s}{r\Hc}+\frac{\Hc'}{\Hc^2}} f + \Upsilon(z)
											\bigg] \delta(\bk, \eta)\, ,\label{Deltarel_k}
\end{align}
where
\begin{equation}\label{eq:eps}
\Upsilon(z) \define \frac{\te-\ga}{1+\ga} \, f - \frac{\ga}{1+\ga} \pa{ \frac{\Hc'}{\Hc^2}f+f^2+\frac{f'}{\Hc} }
\end{equation}
encodes the deviations from Euler's equation. Here we have assumed that the growth of density perturbation, $D$, defined through $\delta(\bk, z)=[D(z)/D_0]\delta(\bk,0),$ is scale-independent. This is true in the quasi-static limit of the models described in sec.~\ref{sec:theory}. In general, deviations from GR can introduce a scale-dependence in $D$, but we will not consider this possibility in the following. The growth rate~$f$, defined through
\begin{equation}
f(z)=\frac{\dd\ln D(a)}{\dd\ln (a)} 
\end{equation}
is therefore scale-independent as well, and so is $\Upsilon(z)$ defined in~\eqref{eq:eps}.

\subsubsection{Expression of the cross-correlation function}

In the cross-correlation of two populations of galaxies, bright~$\B$ and faint~$\F$, the only quantities which depend on the galaxy population in eqs.~\eqref{Deltast_k} and \eqref{Deltarel_k} are the biases $b_\B$ and $b_\F$ and the slopes $s_\B$ and $s_\F$. Following~\cite{Bonvin:2013ogt} we can write, in the flat-sky limit,
\begin{equation}
\label{xiBFall}
\xi\e{BF}(d, \sigma, \bar z)
= \xi\e{BF}\h{st}(d,\sigma, \bar z)+\xi\e{BF}\h{rel}(d, \sigma, \bar z)+\xi\e{BF}\h{wide}(d, \sigma, \bar z)
	+\xi\e{BF}\h{evol}(d,\sigma, \bar z)+\xi\e{BF}\h{lens}(d,\sigma, \bar z)\, ,
\end{equation}
where $\bar z$ denotes the mean redshift of the survey (or of the redshift bin of interest), and we remind the reader that $d$ is the separation between the galaxies, while $\sigma$ denotes the orientation of the pair of galaxies with respect to the mean direction of observation $\bn$.

In eq.~\eqref{xiBFall}, $\xi\e{BF}\h{st}$ is the standard correlation function containing a monopole, quadrupole and hexadecapole in $\sigma$~\cite{Hamilton:1997zq,Kaiser:1987qv}. The relativistic contribution $\xi\e{BF}\h{rel}$ is given by
\begin{align}
\xi\e{BF}\h{rel} &= \frac{\Hc}{\Hc_0} \pa{\frac{D}{D_0}}^2
								\Bigg[(b_\B-b_\F) \, \pa{ \frac{2}{r\Hc} + \frac{\Hc'}{\Hc^2} + \Upsilon(z) }
											+ 3 (s_\F-s_\B) f^2 \pa{ 1-\frac{1}{r\Hc} } \nonumber\\
							&\hspace*{5cm}
											+ 5 (b_\B s_\F-b_\F s_\B) f \pa{ 1-\frac{1}{r\Hc}}
								 			\Bigg] \nu\h{rel}_1(d) \, P_1(\cos\sigma)\nonumber\\
							&\quad
								+ 2\,\frac{\Hc}{\Hc_0}\pa{\frac{D}{D_0} }^2
								(s_\B-s_\F) f \pa{1-\frac{1}{r\Hc}} \nu\h{rel}_3(d) \, P_3(\cos\sigma)\, , \label{xirel}
\end{align}
where $P_\ell$ denotes the Legendre polynomial of degree $\ell$, and
\be
\label{nurel}
\nu\h{rel}_\ell(d)\equiv\frac{1}{2\pi^2}\int \dd k \; k \Hc_0 P_\delta(k)j_\ell(kd),
\qquad\ell=1,3\, .
\ee
Here $P_\delta(k)$ is the matter power spectrum today, $\ev{\delta(\bk)\delta(\bk')}=(2\pi)^3 P_\delta(k)\delta\e{D}(\bk+\bk')$, while $j_\ell$ denotes the spherical Bessel function of degree $\ell$.
The relativistic correlation comes from the correlation between the standard term~\eqref{eq:Delta_standard} and the relativistic term~\eqref{eq:Delta_relativistic} and, as anticipated in sec.~\ref{sec:aymm}, it is completely anti-symmetric, because it consists of a dipole and an octupole in $\sigma$. The correlation of the relativistic contribution with itself also contributes to $\xi\e{BF}\h{rel}$, but we neglect it here, because it only consists of a monopole and a quadrupole, eliminated in the dipole extraction performed in the next paragraph.

The third contribution to eq.~\eqref{xiBFall}, $\xi\e{BF}\h{wide}$, represents the wide-angle corrections to the flat-sky correlation function. As discussed in refs.~\cite{Bonvin:2013ogt, Bonvin:2015kuc} the wide-angle corrections from density and redshift-space distortions also generate a dipole and an octupole in the correlation function. As such, they contaminate the measurement of the relativistic contribution. The wide-angle effects are minimised by using the angle $\sigma$ to extract the dipole, i.e. the angle between the direction of the median of the pair and the direction of observation---see~\cite{Bonvin:2015kuc, Reimberg:2015jma} for a detailed discussion about other possible angles. This contribution is given by
\begin{equation}
\label{xiwide}
\xi\e{BF}\h{wide}=-\frac{2f}{5}(b\e{B}-b\e{F})\,\frac{d}{r}\,\nu\h{st}_2(d)\big[P_1({\cos\sigma})-P_3(\cos\sigma)\big]\, ,
\end{equation}
with
\be
\label{nust}
\nu\h{st}_\ell(d) \equiv \frac{1}{2\pi^2}\int \dd k  \; k^2 P_\delta(k)j_\ell(kd),
\qquad\ell=0,2,4\, .
\ee
Comparing eq.~\eqref{xiwide} with eq.~\eqref{xirel}, we see that $\xi\e{BF}\h{wide}$ is suppressed by $d/r\ll 1$ with respect to $\xi\e{BF}\h{rel}$. However $\nu\h{st}_\ell$ is enhanced by a factor $k/\Hc\gg 1$ with respect to $\nu\h{rel}_\ell$. Since $d/r\sim \Hc/k$ these two effects compensate and the wide-angle correction becomes of the same order of magnitude as the relativistic contribution (see also fig. 8 of~\cite{Bonvin:2013ogt}). 

The fourth term in eq.~\eqref{xiBFall}, $\xi\e{BF}\h{evol}$, denotes the evolution corrections. These are due to the fact that the biases, growth rate, and slopes of the luminosity function evolve with redshift, and consequently also gives rise to a dipole and octupole. However, as shown in fig.~11 of~\cite{Bonvin:2013ogt}, these evolution corrections are always sub-dominant with respect to both the relativistic contribution and the wide-angle corrections. We can therefore safely neglect them.

Finally, the last term in eq.~\eqref{xiBFall}, $\xi\e{BF}\h{lens}$, represents the lensing contribution. Again, the lensing contribution to the dipole and octupole is significantly smaller than the relativistic and wide-angle contributions, and hence we neglect it---see fig.~8 of~\cite{Bonvin:2013ogt}.

\subsubsection{Estimator of the dipole}

From eqs.~\eqref{xiBFall} and~\eqref{xirel} we see that the optimal way to test Euler's equation is to extract the dipolar modulation in the two-point correlation function. First, this allows us to completely get rid of the standard correlation function $\xi\e{BF}\h{st}$ (which is insensitive to the parameters $\Theta$ and $\Gamma$ that we want to constrain). And second, it strongly reduces the impact of the evolution and lensing contributions which are negligible in the dipole. Hence the only two relevant contributions to the dipole are the wide-angle contribution and the relativistic contribution. As shown in eq.~\eqref{xiwide} the wide-angle contribution is insensitive to the parameters $\Theta$ and $\Gamma$. As such this contribution is a contamination that we would like to remove.  As discussed in~\cite{Bonvin:2013ogt}, this can be done observationally by removing the quadrupole of the bright and faint populations from the dipole, i.e. by constructing the following estimator
\begin{align}
\hat{\xi}^{1}\e{BF}=&\frac{3}{8\pi}\left(\frac{\ell\e{p}}{d}\right)^2\frac{\mathcal{V}}{\ell\e{p}^3}\sum_{ij}
\big[\Delta_{\B}(\bx_i)\Delta_{\F}(\bx_j)-\Delta_{\F}(\bx_i)\Delta_{\B}(\bx_j)\big]P_1(\cos\sigma_{ij})\delta\e{K}(d_{ij}-d)\label{estimator}\\
&+\frac{3}{10}\frac{d}{r}\frac{5}{4\pi}\left(\frac{\ell\e{p}}{d}\right)^2\frac{\mathcal{V}}{\ell\e{p}^3}\sum_{ij}
\big[\Delta_{\B}(\bx_i)\Delta_{\B}(\bx_j)-\Delta_{\F}(\bx_i)\Delta_{\F}(\bx_j)\big]P_2(\cos\sigma_{ij})\delta\e{K}(d_{ij}-d)\nonumber\, ,
\end{align}
where $\mathcal{V}$ denotes the volume of the survey (or of the redshift bin of interest), $\ell\e{p}$ is the length of the cubic pixel in which $\Delta$ is measured, the sum runs over all pairs of pixels $i, j$ in the survey and $\delta\e{K}$ is the Kronecker delta function. The first line in~\eqref{estimator} isolates the dipole contribution in the cross-correlation between bright and faint galaxies (the minus sign ensures that the correlation does not vanish under the exchange of bright and faint galaxies). The second line removes the wide-angle effect. Taking the mean of the estimator and going to the continuous limit\footnote{The continuous limit is obtained by replacing the sum over pixels by a 3-dimensional integral $\sum_i\rightarrow\frac{1}{\ell\e{p}^3}\int \dd^3 x$ and the Kronecker delta function by a Dirac delta function $\delta\e{K}(d_{ij}-d)\rightarrow \ell\e{p}\delta\e{D}(|\bx-\by|-d)$.} we find indeed
\begin{align}
\ev{\hat{\xi}^{1}\e{BF}}
=& \frac{3}{2}\int_{-1}^1\dd\mu \; \pac{\xi\e{BF}\h{rel}(d, \mu, \bar z) + \xi\e{BF}\h{wide}(d, \mu, \bar z)} P_1(\mu)
	\nonumber\\
&\quad + \frac{3}{10} \frac{d}{r} \frac{5}{2} \int_{-1}^1 \dd\mu 
				\pac{ \xi\e{BB}\h{st}(d, \mu, \bar z)-\xi\e{FF}\h{st}(d, \mu, \bar z)} P_2(\mu)
\nonumber\\
=& \frac{\Hc}{\Hc_0} \pa{\frac{D}{D_0}}^2
		\Bigg[ (b_\B-b_\F) \pa{ \frac{2}{r\Hc} + \frac{\Hc'}{\Hc^2} + \Upsilon(z) }
\nonumber\\
&\quad + 3 (s_\F-s_\B)f^2 \pa{ 1-\frac{1}{r\Hc}} + 5  (b_\B s_\F-b_\F s_\B) f \pa{1-\frac{1}{r\Hc}}
				\Bigg]\nu\h{rel}_1(d)\, , \label{estimator_final}
\end{align}
with $\mu\define\cos\sigma$. Note that eq.~\eqref{estimator_final} is valid in the linear regime. An equivalent estimator has been constructed fo measure gravitational redshift in clusters~\cite{Wojtak:2011ia,Sadeh:2014rya} and in the non-linear regime of large-scale structure~\cite{Alam:2017izi}. The modelling and interpretation of the signal in the non-linear regime is however much more complicated~\cite{Zhao:2012st,Kaiser:2013ipa,Cai:2016ors} and its use to test Euler's equation is not straightforward. 

\section{Forecasts: are deviations from Euler detectable?}
\label{sec:forecasts}

We now forecast the detectability of deviations from Euler's equation in future surveys. We consider first the SKA survey and then the DESI survey.

\subsection{Constraints from the SKA}

In its second phase of operation, the SKA will observe $8.8\times 10^8$ galaxies over 30'000 square degrees, from redshift 0.1 to redshift 2. We split the redshift range into bins of width $\Delta z=0.1$ and use the number densities derived in table 3 of~\cite{Bull:2015lja}. In each redshift bin, we calculate the signal $\ev{\hat{\xi}^{1}\e{BF}}$ from eq.~\eqref{estimator_final} and the covariance matrix $C^1_{\B\F}$. The covariance matrix has been calculated in~\cite{Bonvin:2015kuc,Hall:2016bmm}. In general, covariance matrices contain a shot noise contribution, a cosmic variance contribution and a mixed contribution. However, as shown in~\cite{Bonvin:2015kuc,Hall:2016bmm}, the dipole estimator automatically removes the cosmic variance contribution from density and redshift-space distortions. Fitting for a dipole in the data allows us therefore not only to get rid of the dominant standard contribution (density and redshift-space distortion) in the \emph{signal} but also in the \emph{noise}. As such, the dipole estimator is a powerful tool to isolate the relativistic contributions. The expression for the covariance can be found in appendix~\ref{app:cov}. Note that subtracting the wide-angle effects adds an additional contribution to the covariance. However, as shown in~\cite{Hall:2016bmm}, this contribution is always strongly subdominant with respect to the shot noise and mixed contribution (see fig. 2 of~\cite{Hall:2016bmm}).

We split the population of galaxies into a bright and a faint population with the same number of galaxies. We use the mean redshift given in table 3 of~\cite{Bull:2015lja} and we assume that the two populations are equally distributed around this mean redshift, with a redshift difference $b_\B-b_\F=0.5$. The signal-to-noise is directly proportional to the bias difference. As an example, in BOSS, a bias difference of 1 has been measured between the bright and faint populations of luminous red galaxies (LRG's)~\cite{Gaztanaga:2015jrs}. In the main galaxy sample of SDSS, galaxies have been split into six populations according to their luminosity, with a bias ranging from 0.96 to 2.16~\cite{Percival:2006gt,Cresswell:2008aa}. For the HI galaxies targeted by the SKA, the expected bias difference is less well known, but a difference of 0.5 seems quite conservative. As we will see in sec.~\ref{sec:SKA_mult} the constraints on deviations in Euler's equation scale directly with the bias difference. The signal-to-noise and the constraints depend also on the slope of the luminosity function $s_\B$ and $s_\F$. We fix those to be zero. We assume a background cosmology consistent with $\Lambda$CDM, with the fiducial cosmological parameters from BOSS DR11~\cite{Anderson:2013zyy}: $\Omega_{\rm m}=0.274, h=0.7, \Omega_{\rm b} h^2=0.224, n_{\rm s}=0.95$ and $\sigma_8=0.8$. 

\begin{figure}[!t]
\centering
\includegraphics[width=0.48\columnwidth]{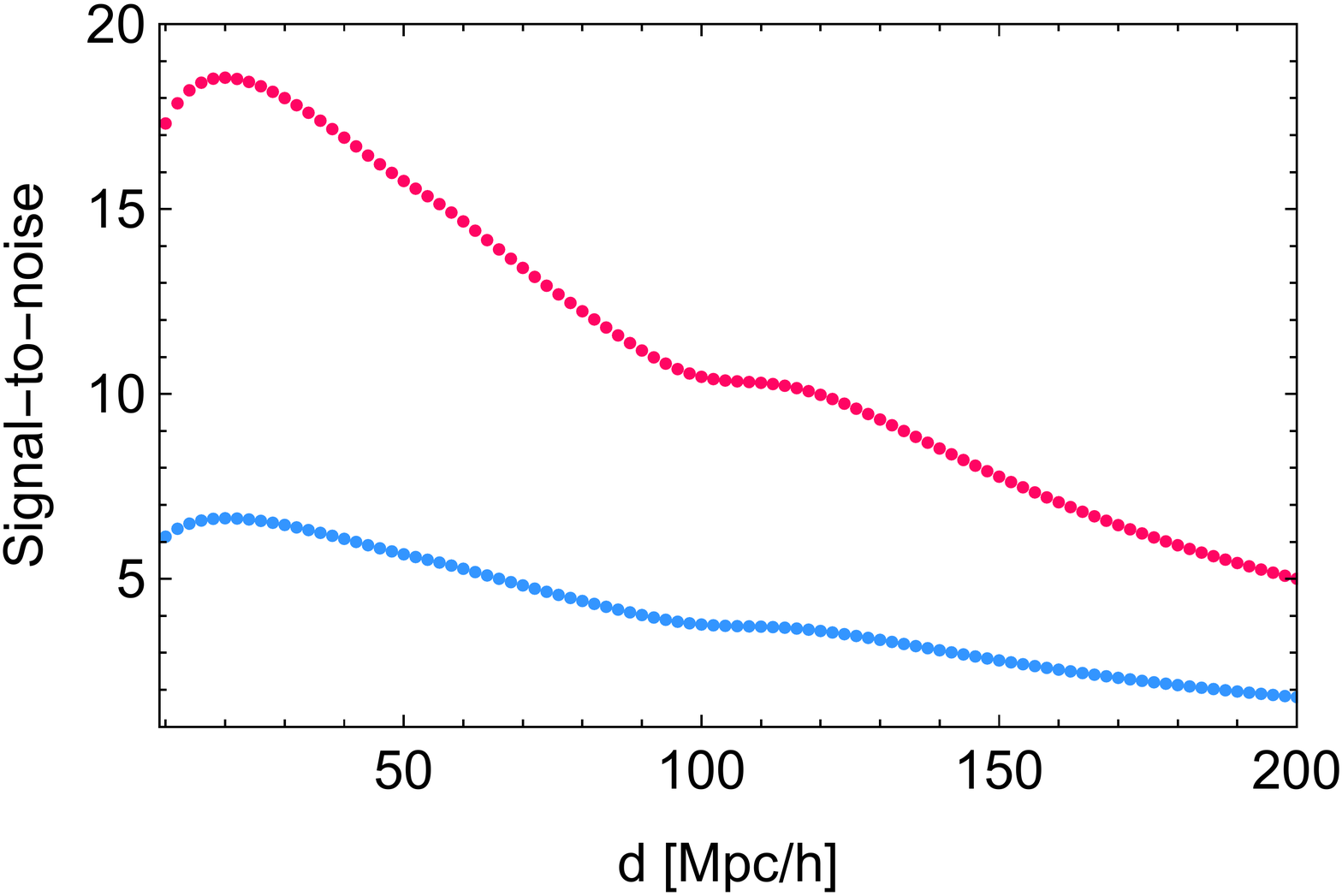}\hspace{0.5cm}\includegraphics[width=0.48\columnwidth]{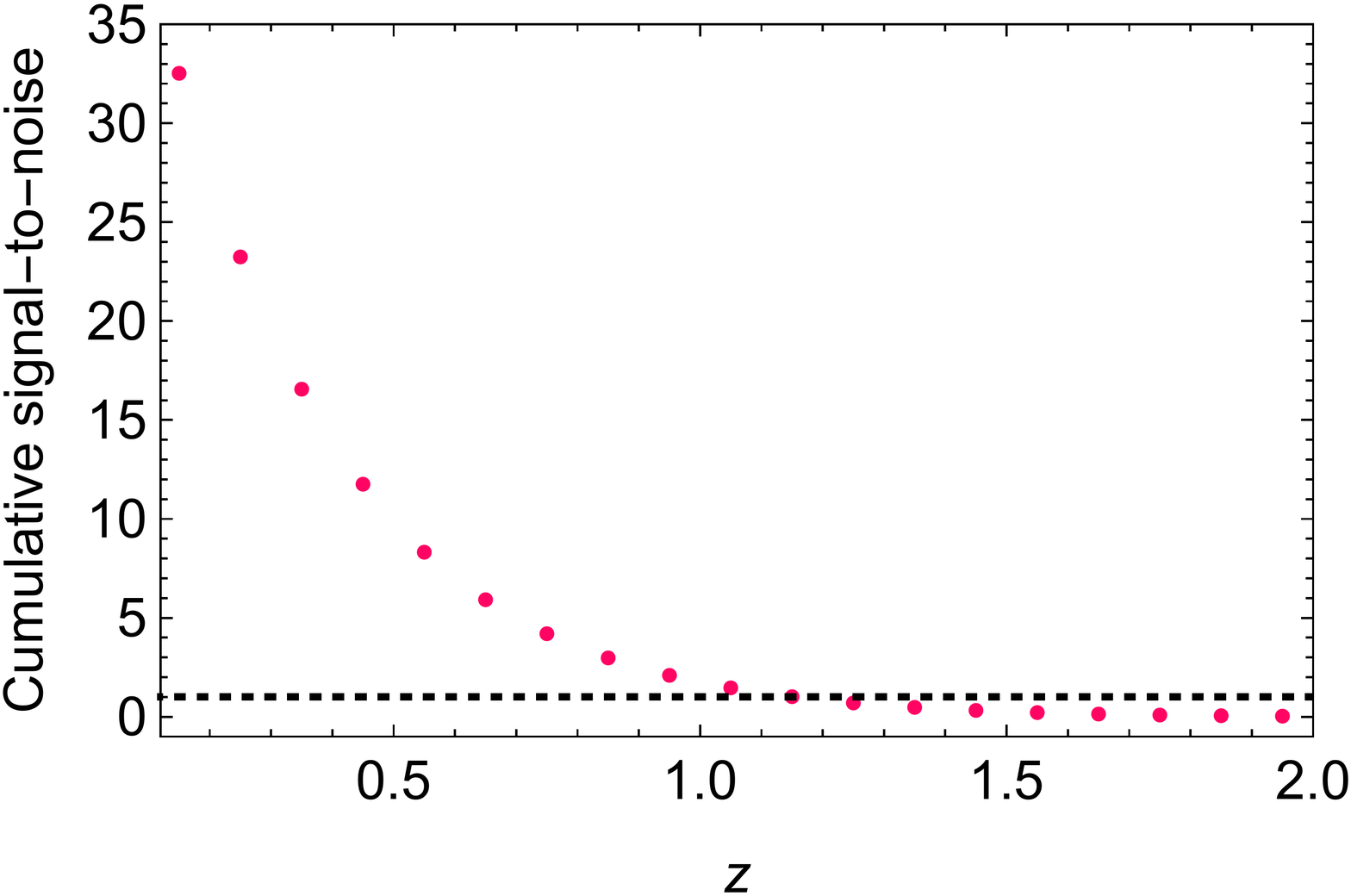}
\caption{\emph{Left panel}: Signal-to-noise of the dipole in the SKA, plotted as a function of separation $d$, with pixel size $\ell_{\rm p}=2\mph$. The red dots show the signal-to-noise at $\bar z=0.15$ and the blue dots at $\bar z=0.45$. \emph{Right panel}: the cumulative signal-to-noise from $10\leq d\leq 200\mph$, per bin of redshift from $0.1\leq z\leq 2$.}
\label{fig:SN_dipole}
\end{figure}

In fig.~\ref{fig:SN_dipole} we plot the signal-to-noise of the dipole. In the left panel, we show the signal-to-noise at each separation $d$, in the lowest redshift bin $0.1\leq z \leq 0.2$ (red dots) and in the bin $0.4\leq z\leq 0.5$ (blue dots). We use a pixel size $\ell_{\rm p}=2\mph$. We see that the signal-to-noise peaks around $20\mph$. In the right panel we show the cumulative signal-to-noise from $10\leq d\leq 200\mph$, as a function of redshift. Above $z=1.2$ the cumulative signal-to-noise drops below 1, meaning that the dipole is not observable. The cumulative signal-to-noise over the whole range of separation and redshift is of 46.4, showing that the dipole will be robustly measured with the SKA.

\subsubsection{Constraints from the dipole}
\label{sec:forecast_dipole}

We first forecast the constraints on $\Theta$ and $\Gamma$ that we would obtain using only the dipole estimator defined in eq.~\eqref{estimator}. We fix all cosmological parameters and the biases to their fiducial value. Cosmological parameters are indeed well measured by the CMB. The biases on the other hand are unconstrained by the CMB, but as we will see in sec.~\ref{sec:SKA_mult} they can be robustly constrained by the standard multipoles.
For $\Theta$ and $\Gamma$ we assume that they evolve according to
\be
\label{evol}
\Theta(z)=\frac{1-\Omega_{\rm m}(z)}{1-\Omega_{\rm m0}}\Theta_0\,,
\ee
where $\Theta_0\equiv\Theta(z=0)$, and similarly for $\Gamma(z)$. As discussed in~\cite{Gleyzes:2015rua} this evolution ensures that the deviations vanish when the background dark energy density is negligible, like at high redshift, and one recovers Euler's equation. We use fiducial values $\Theta_0=\Gamma_0=0$.

As a first example we assume that $\Theta$ and $\Gamma$ are the only deviations from general relativity, in particular we assume that the growth function $D(z)$ evolves as in a $\Lambda$CDM Universe. In the scalar-tensor and vector-tensor models described in sec.~\ref{sec:theory} deviations from Euler's equations are usually accompanied by deviations in the growth rate. However one could imagine other models where this is not the case, and the only deviations from general relativity would be in Euler's equation. 

We construct the Fisher matrix for $\Theta_0$ and $\Gamma_0$
\be
\label{Fisherdip}
F_{ab}=\sum_{\bar z}\sum_{ij}\frac{\partial\ev{\hat{\xi}^{1}\e{BF}(d_i, \bar z)}}{\partial p_a}\left[C^1_{\B\F}\right]^{-1}(d_i,d_j, \bar z)\frac{\partial\ev{\hat{\xi}^{1}\e{BF}(d_j, \bar z)}}{\partial p_b}\, ,
\ee
where $p_a=\Theta_0,\Gamma_0$, and $C^1_{\B\F}$ denotes the covariance matrix of the dipole estimator. 
In~\eqref{Fisherdip}, the sum runs over all pixel's separations $d_i$ and all redshift bins. We account for correlations between separations (see appendix~\ref{app:cov}), but we neglect correlations between different redshift bins. We use pixel's size of $\ell_{\rm p}=2\mph$  and pixel's separations $d_{\rm min}\leq d_i\leq d_{\rm max}$. We choose $d_{\rm max}=200\mph$ since we have checked that the constraints do not improve if we include larger separations. For the minimum separation, we use $d_{\rm min}=10\mph$. At this scale the impact of non-linearities on the dipole is of the order of 10\,\%\footnote{We estimate the impact of non-linearities on the dipole in the following way: we use the linear continuity equation to express the velocity in terms of the density, and then we use halo-fit to compute the non-linear density. This procedure is of course not completely correct since non-linearities modify the continuity equation, but it allows us to estimate the scale at which non-linearities become important.}. As we will see below, increasing $d_{\rm min}$ to $20\mph$, where the impact of non-linearities on the dipole is already less than 2\,\%, has almost no effect on the constraints on $\Theta_0$ and $\Gamma_0$.  

\begin{figure}[!t]
\centering
\includegraphics[width=0.54\columnwidth]{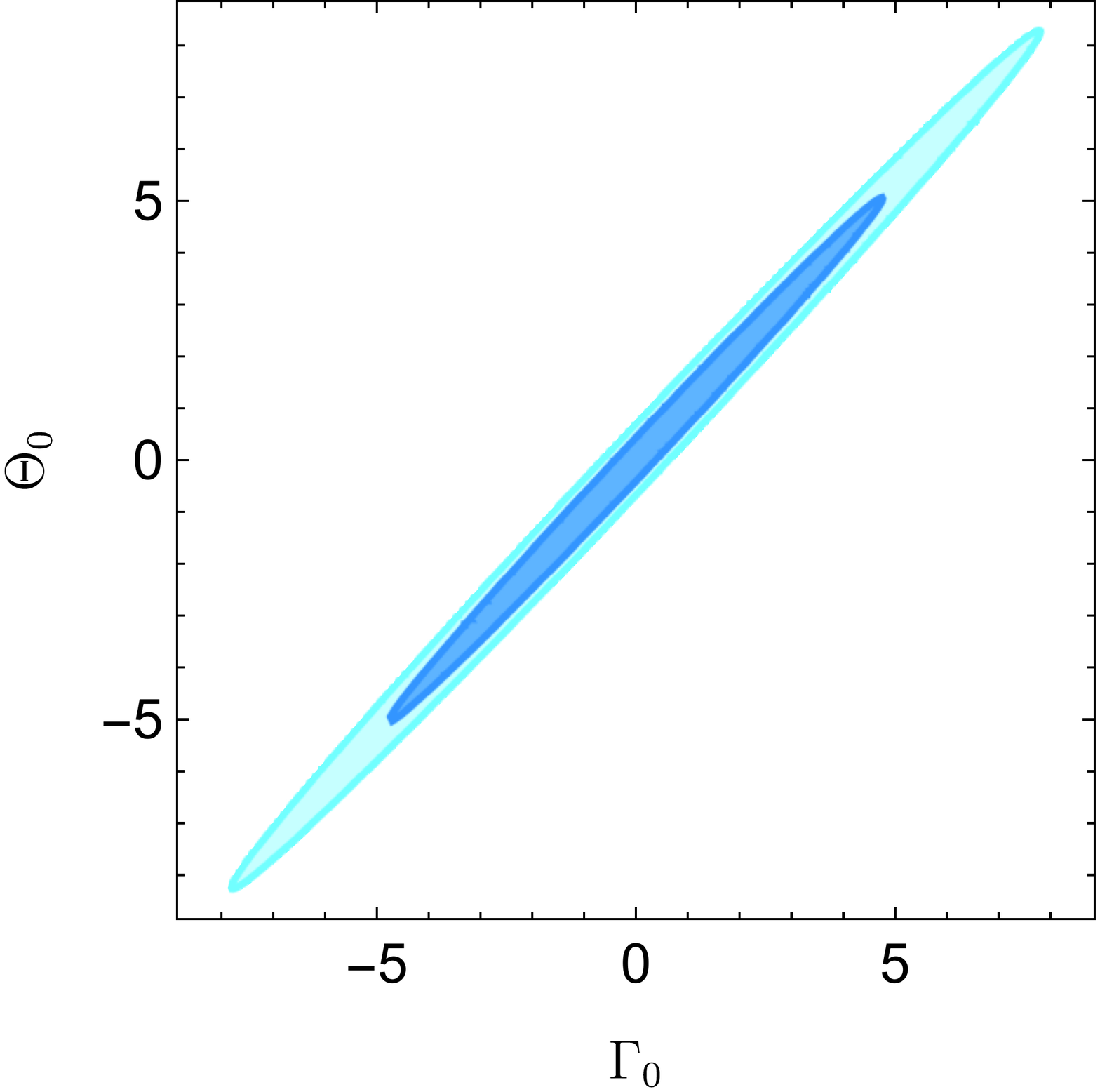}\hspace{-1.7cm}\includegraphics[width=0.54\columnwidth]{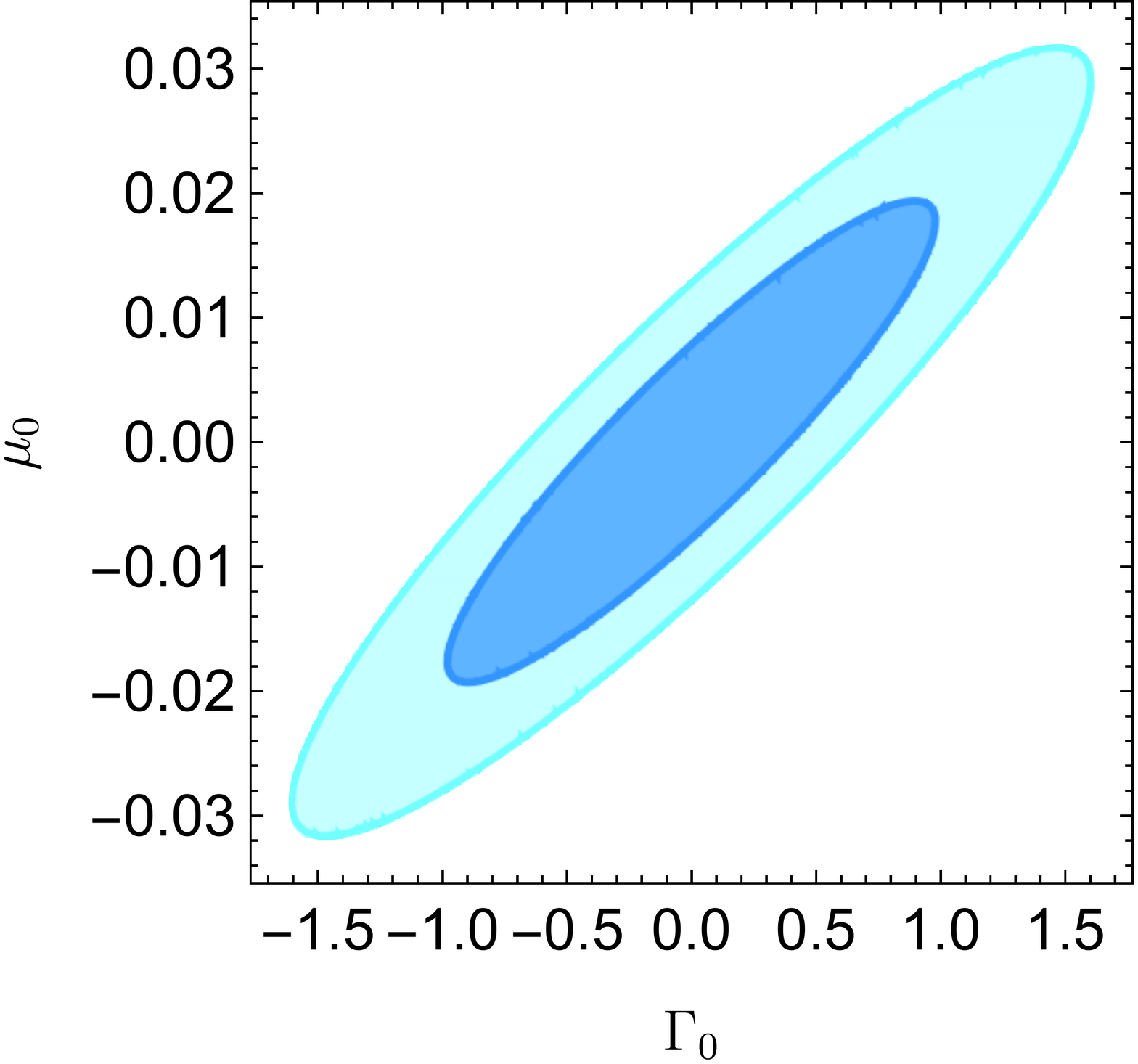}
\caption{\emph{Left panel}: Joint 1-$\sigma$ and 2-$\sigma$ constraints on the parameters $\Theta_0$ and $\Gamma_0$ obtained from the dipole only, with the SKA. The other cosmological parameters and the biases are fixed to their fiducial value and the growth rate is the one of a $\Lambda$CDM universe: $\mu_0=0$.  
\emph{Right panel}: Joint 1-$\sigma$ and 2-$\sigma$ constraints on the parameters $\Gamma_0$ and $\mu_0$ obtained from the dipole only, with the SKA. The other cosmological parameters and the biases are fixed to their fiducial value and $\Theta_0=0$.}
\label{fig:dipole_constraints}
\end{figure}

In the left panel of fig.~\ref{fig:dipole_constraints} we show the joint constraints on $\Theta_0$ and $\Gamma_0$. There is a strong degeneracy between the two parameters, leading to relatively large marginalised 1-$\sigma$ errors of $\Delta{\Theta_0}=3.33$ and $\Delta{\Gamma_0}=3.14$. This degeneracy can be understood from eq.~\eqref{estimator_final}. We see that the amplitude of the dipole depends on $\Theta_0-\Gamma_0$ [first term in the expression~\eqref{eq:eps} of $\Upsilon$] as well as on $\Gamma_0$ individually. However this second dependence is much weaker than the first one at low redshift, since the terms in the second parenthesis are proportional to the second time derivative of the growth factor, $D''$. As a consequence $\Theta_0$ and $\Gamma_0$ are strongly correlated. However, as discussed in sec.~\ref{sec:parametrisation}, in vector-tensor theories, $\Theta=0$. In the EFT of dark energy $\Theta$ is generically not zero, but it is directly proportional to the background evolution of the scalar field. As such it is generically expected to be significantly smaller than $\Gamma$. It is therefore natural to explore how the constraints change when $\Theta=0$. In this case, the constraints on $\Gamma_0$ strongly tightens and we obtain $\Delta{\Gamma_0}=0.26$.

Comparing with the constraints on standard modified gravity parameters in~\cite{Bellini:2015xja, Alonso:2016suf} we see that the constraints on $\Gamma_0$ are weaker by 1 to 3 orders of magnitude. This is not surprising since the signal-to-noise of the dipole is weaker than that of the standard multipoles. However, as discussed in the introduction, a constraint on $\Gamma_0$ would provide the first direct test of the equivalence principle at cosmological scale, which cannot be probed in a model-independent way with the standard multipoles. In~\cite{Gleyzes:2015rua}, constraints on Euler's equation are obtained indirectly in the specific case of scalar-tensor theories. In this model, the parameters that modify Euler's equation also modify the growth of structure, which is measured via the standard multipoles. These constraints are therefore directly related to the specific choice of model: scalar-tensor theories. Constraints of the order of $10^{-3}$ are obtained in this case, varying each parameter individually. Marginalising over all modified gravity parameters degrade the constraints significantly~\cite{Leung:2016xli}. This shows the complementarity of relativistic effects, which break the degeneracy between parameters and which can directly probe Euler's equation, without underlying assumptions about the model.

\begin{figure}[!t]
\centering
\includegraphics[width=0.49\columnwidth]{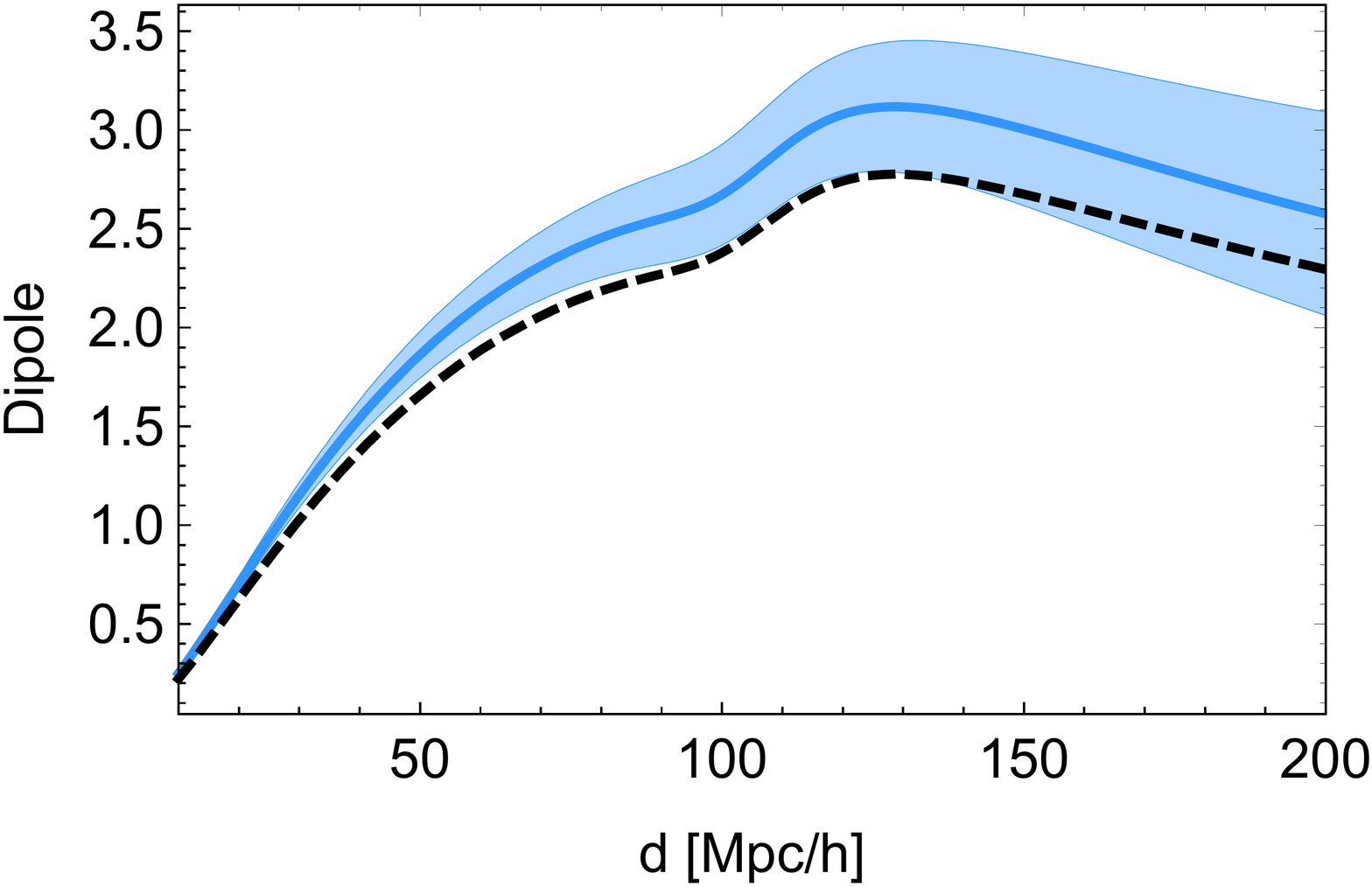}\hspace{0.2cm}\includegraphics[width=0.49\columnwidth]{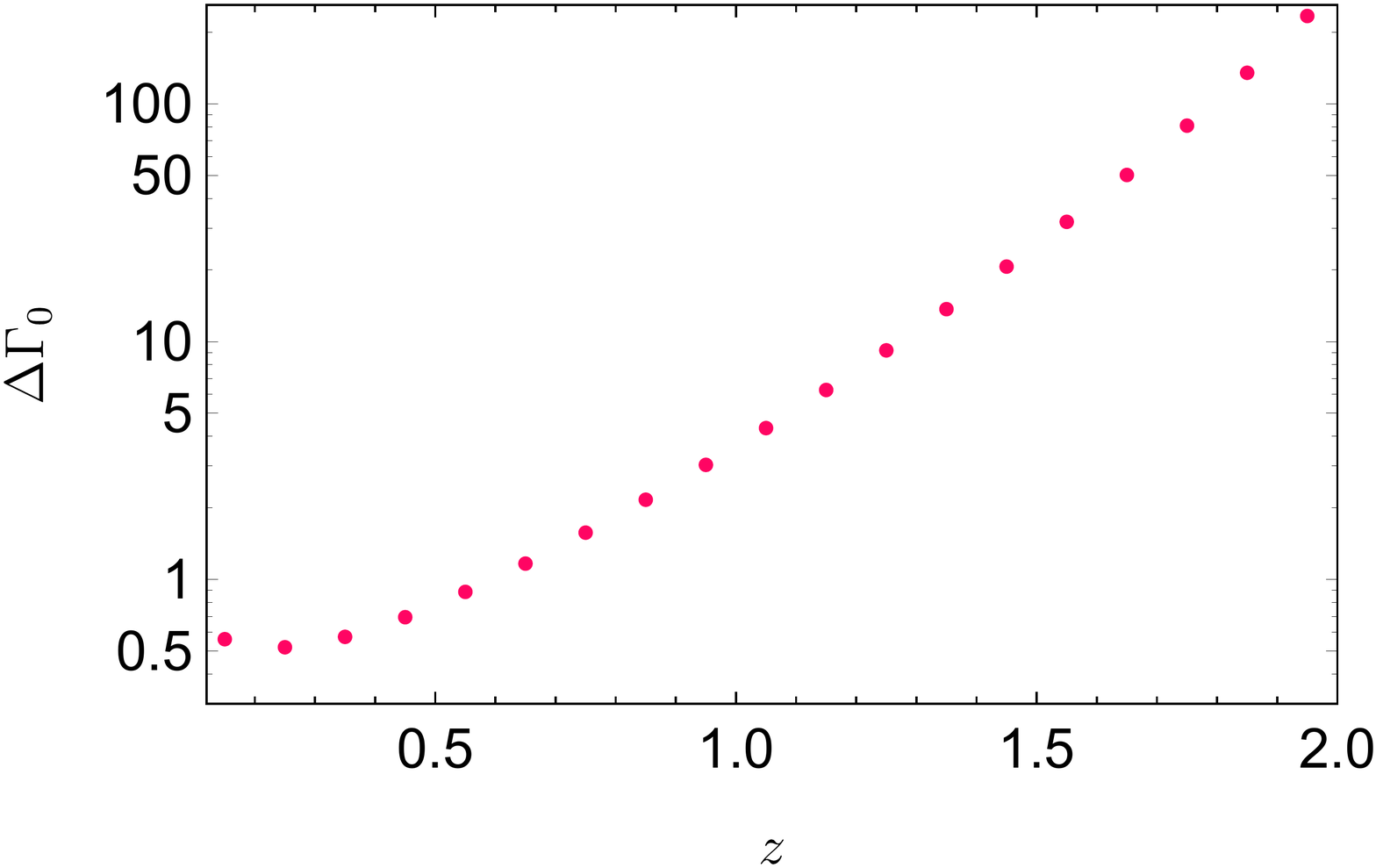}
\caption{\emph{Left panel}: The dipole (multiplied by $d^2$), plotted as a function of separation $d$ in the lowest redshift bin of the SKA: $0.1\leq z\leq 0.2$. The blue solid line shows the dipole in a $\Lambda$CDM universe, with $\Gamma_0=\Theta_0=0$. The dashed region represent the error bars calculated from eq.~\eqref{cov1}. The dashed black line shows the dipole when the EEP is violated, with $\Gamma_0=2$ and $\Theta_0=0$. \emph{Right panel}: The constraints on $\Gamma_0$, keeping all other parameters fixed to their fiducial value, in each redshift bin of the SKA.}
\label{fig:dipole_dev}
\end{figure}

In fig.~\ref{fig:dipole_dev} (left panel) we show the dipole in the lowest redshift bin $\bar z=0.15$ when $\Gamma_0=0$ (blue solid line) and when $\Gamma_0=2$ (dashed black line). We see that in this case the deviation due to the breaking of the equivalence principle is larger than the error bars (coloured blue region) in the range $10\leq d\leq 130\mph$. Combining all separations and all redshifts allow us to see deviations already for $\Gamma_0=0.26$. In the right panel of fig.~\ref{fig:dipole_dev}, we show the constraints on $\Gamma_0$ in each redshift bin. We see that the constraints come mainly from low redshift.

As discussed above, generic modified gravity models generate not only deviations in Euler's equation, but also deviations in the way structure are growing as a function of time, i.e. in the growth function $D$. We model this in the following way
\be
\label{D1mod}
D(z)=\bar D(z)\big[1+\mu(z) \big]\, ,
\ee
where $\bar D$ denotes the growth function in a $\Lambda$CDM universe, and we let the deviation $\mu(z)$ evolve as in eq.~\eqref{evol}. The growth rate can then be written as a function of $\mu_0$
\be
\label{fmod}
f(z)=\bar f(z)+3\Omega_{{\rm m}0}(1+z)^3\left[\frac{1-\Omega_{\rm m}(z)}{1-\Omega_{\rm m0}}\right]^2\mu_0\, ,
\ee
where $\bar f$ denotes the growth rate in a $\Lambda$CDM universe. Inserting~\eqref{D1mod} and \eqref{fmod} into eq.~\eqref{estimator_final}, and linearising in the deviations, we can forecast the constraints on $\Gamma_0$ and $\mu_0$, fixing $\Theta=0$. 

The joint constraints are shown in the right panel of fig.~\ref{fig:dipole_constraints}. We see that $\Gamma$ is less degenerated with $\mu$ than with $\Theta$. This comes from the different time dependence that multiplies these two parameters in eq.~\eqref{estimator_final}. The marginalised constraints on each of the parameters are given by $\Delta{\Gamma_0}=0.65$ and $\Delta{\mu_0}=0.013$. Allowing for the growth rate to vary degrades therefore the constraints on $\Gamma_0$ by a factor 2.5. 

Finally, let us note that all these constraints are obtained while fixing the bias of the bright and faint populations to their fiducial value. If instead we let the biases vary, we loose all constraining power, since the bias difference in each redshift bin is strongly degenerated with a change in the growth rate  and with modifications to Euler's equation. However, in addition to measuring the dipole of the correlation function, one can also measure the monopole, quadrupole and hexadecapole of the bright and faint populations separately. As we will see in the next section, this allows us to break the degeneracy between the biases, the growth rate $\mu_0$ and the modification to Euler's equation $\Gamma_0$. 

\subsubsection{Constraints from the dipole combined with standard multipoles}
\label{sec:SKA_mult}

We now consider 7 observables: the cross-correlation dipole, and the monopole, quadrupole and hexadecapole of the bright and of the faint populations. In the forecasts, we neglect the covariance between different multipoles. We take however into account the covariance between the monopole of the bright and of the faint population, as well as between their quadrupole and hexadecapole. These covariance matrices have been calculated in~\cite{Hall:2016bmm} and are summarised in appendix~\ref{app:cov}. 

As in the previous section, we fix the cosmological parameters to their $\Lambda$CDM fiducial value. We consider $\Gamma_0$ and $\mu_0$ as free parameters (as motivated above we fix $\Theta=0$). We assume that the bias of the bright and faint population evolves in the following way~\cite{Bull:2015lja}
\bea
b_\B(z)&=&b_{1} e^{b_{2} z}+\frac{\Delta b}{2}\, ,\\
b_\F(z)&=&b_3 e^{b_4 z}-\frac{\Delta b}{2}\, ,
\eea
where as before we fix the bias difference $\Delta b=b_\B-b_\F=0.5$. We have therefore four additional free parameters $b_1, b_2, b_3$ and $b_4$ that we want to constrain. We choose their fiducial value as in~\cite{Bull:2015lja} (see table 6): $b_1=b_3=0.554$ and $b_2=b_4=0.783$. In this way the bias of the bright and of the faint population are uncorrelated, but their difference is always of 0.5. As shown below, the constraints can easily be rescaled according to the bias difference.

\begin{figure}[!t]
\centering
\includegraphics[width=0.65\columnwidth]{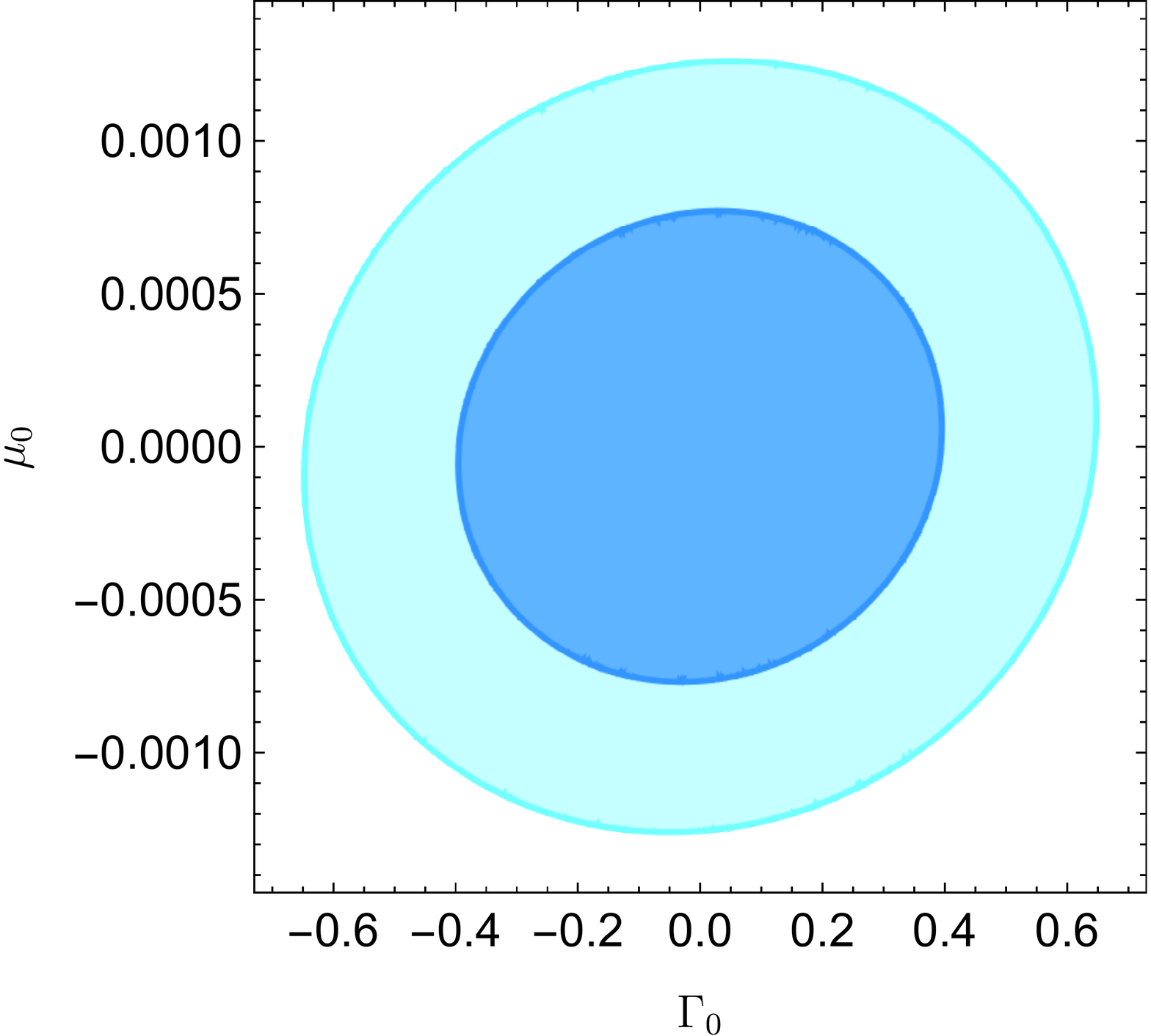}
\caption{Joint 1-$\sigma$ and 2-$\sigma$ constraints on the parameters $\Gamma_0$ and $\mu_0$ obtained from a combination of the dipole, the monopole, quadrupole and hexadecapole of the bright and faint population, with the SKA. Here $\Theta_0=0$, and the constraints are marginalised over the four bias parameters, $b_1$ to $b_4$.}
\label{fig:combined_constraints}
\end{figure} 

In fig.~\ref{fig:combined_constraints} we show the joint constraints on $\Gamma_0$ and $\mu_0$ marginalised over the four bias parameters. We see that adding the standard multipoles strongly tightens the constraints on $\mu_0$: the marginalised error on $\mu_0$ becomes $\Delta\mu_0=5.1\times 10^{-4}$, consistent with~\cite{Gleyzes:2014rba}. This is not surprising since the monopole, quadrupole and hexadecapole provide three different combinations of $\mu_0$ and of the bias, which can then be both robustly constrained. Since the signal-to-noise of the even multipoles is significantly larger than that of the dipole, we find that adding the dipole in the forecasts \emph{does not} improve the constraints on $\mu_0$ at all. This is consistent with the conclusion of~\cite{Lorenz:2017iez} which showed (using the angular power spectrum $C_\ell$) that relativistic effects do not improve the constraints on parameters that govern the growth of structure, like $\mu_0$ (see however~\cite{Lombriser:2013aj} for specific cases, where this may not be true). The usefulness of relativistic effects is therefore to test for deviations that {\it cannot} be probed with standard observables, like deviations in the equivalence principle. The marginalised error on $\Gamma_0$ is also tightened by the combined analysis: $\Delta\Gamma_0=0.26$. Hence even though the standard multipoles are not sensitive to $\Gamma_0$, they improve the constraints on $\Gamma_0$ by breaking the degeneracy with $\mu_0$. Comparing with the results in sec.~\ref{sec:forecast_dipole}, we see that we recover the constraints that we had on $\Gamma_0$ from the dipole  when all other parameters were kept fixed. Combining the dipole alone with the standard multipoles of the bright and faint populations provides therefore an ideal way of testing for deviations in both the growth of structure $\mu_0$ and in Euler's equation $\Gamma_0$. 

Note that in these forecasts we have assumed that $\Gamma$ evolves as in eq.~\eqref{evol}. If instead we choose a constant $\Gamma$ over the whole redshift range (in the model of~\cite{Gleyzes:2014rba}, this would correspond to a fixed coupling of the dark matter to the scalar field), then the constraints on $\Gamma$ tightens to $\Delta\Gamma=0.17$. This comes from the fact that in this case high redshift bins contribute more to the constraints since $\Gamma$ does not decrease with redshift.

For these forecasts we have used as minimum separation $d_{\rm min}=10\mph$. Increasing this to $20\mph$, in order to reduce the impact of non-linearities, does very slightly degrade the constraint on $\Gamma_0$ from $\Delta\Gamma_0=0.26$ to $\Delta\Gamma_0=0.29$. The constraint on $\mu_0$ is more sensitive to $d_{\rm min}$ since it changes from $\Delta\mu_0=5.1\times 10^{-4}$ to $\Delta\mu_0=1.2\times 10^{-3}$. This can be understood by the fact that the standard multipoles are more strongly affected by small scales than the dipole. Comparing eq.~\eqref{nurel} with eq.~\eqref{nust} we see indeed that the standard multipoles contain a factor $k/\Hc_0$ more than the relativistic dipole. As such, at a given separation, the dipole is less sensitive to large $k$'s than the standard multipoles.  

Finally, as discussed before, the forecasts are directly sensitive to the bias difference between the bright and faint population, since the estimator~\eqref{estimator_final} is proportional to $b_\B-b_\F$. Here we have used a fixed difference: $b_\B-b_\F=0.5$. Increasing this bias difference to 1, we find that the constraint on $\Gamma_0$ tightens by roughly a factor 2, $\Delta\Gamma_0=0.14$, while the constraint on $\mu_0$ remains almost the same, $\Delta \mu_0=5\times 10^{-4}$. Equivalently decreasing the bias difference to 0.25 decreases the constraint on $\Gamma_0$ by roughly a factor 2, $\Delta\Gamma_0=0.51$, while the constraint on $\mu_0$ remains almost the same, $\Delta \mu_0=5.3\times 10^{-4}$. Hence as expected the constraint on $\Gamma_0$ scales directly with the bias difference $b_\B-b_\F$. 

\subsection{Constraints from DESI}

In the previous section we have seen that the SKA will provide meaningful constraints on deviations from the equivalence principle. On a shorter timescale, the DESI survey~\cite{Aghamousa:2016zmz} is expected to map galaxies over a large range of redshifts and scales and could already deliver interesting tests of the equivalence principle. The DESI survey will observe different types of galaxies at different redshifts. At low redshift, $0\leq z\leq 0.5$, the Bright Galaxy Sample (BGS) is expected to contain approximately 10 million of galaxies. At middle redshift, $0.6\leq z\leq 1$, DESI will observe Luminous Red Galaxies (LRG) and Emission Line Galaxies (ELG). Finally at high redshift, $1\leq z\leq 1.7$, only ELG's will be detected. The area of the survey is expected to reach 14'000 square degrees. For our forecasts, we use the bias evolution given in sec. 3 of~\cite{Aghamousa:2016zmz} and the number densities and volumes given in tables 2.3 and 2.5. 
We forecast the constraints on $\Gamma_0, \mu_0$, and the biases in each range of redshifts, and we then combine them. As before we neglect the covariance between individual redshift bins. 

In the lowest range of redshift, $0\leq z\leq 0.5$, we split the 10 millions BGS into two equal populations (bright and faint) with a bias difference $\Delta b=0.5$. We assume that the bias of each population evolves as 
\bea
b_\B(z)&=&b_1\frac{D_0}{D(z)}+\frac{\Delta b}{2}\, ,\\
b_\F(z)&=&b_2\frac{D_0}{D(z)}-\frac{\Delta b}{2}\, ,
\eea
where $b_1$ and $b_2$ are two free parameters with fiducial value $b_1=b_2=1.34$~\cite{Aghamousa:2016zmz}.

In the middle range of redshift, $0.6\leq z\leq 1$, we have two populations of galaxies, the LRG's and the ELG's, with very different biases. However, the number density of ELG's is significantly higher than that of LRG's. Hence it is not optimal to use only cross-correlations between these two populations. We split therefore the galaxies into three populations. Population A contains all the LRG's with a bias
\be
b_{\rm A}=b_3\frac{D_0}{D(z)}\, ,
\ee
where the fiducial value  $ b_3=1.7$. Population B contains half of the ELG's and population C the other half, with biases
\bea
b_\B(z)&=&b_4\frac{D_0}{D(z)}+\frac{\Delta b}{2}\, ,\\
b_{\rm C}(z)&=&b_5\frac{D_0}{D(z)}-\frac{\Delta b}{2}\, ,
\eea
where $\Delta b=0.5$ is fixed and the fiducial values of the free parameters is $b_4=b_5=0.84$~\cite{Aghamousa:2016zmz}. With this split we have a similar number of galaxies in each population:  3.9 millions LRG's, 4.6 millions bright ELG's and 4.6 millions faint ELG's. As shown in~\cite{Bonvin:2015kuc}, we can then increase the signal-to-noise of the dipole estimator by weighting each pair of galaxies by the bias difference between the populations. This weighting is optimal in the regime where shot noise dominates. The covariance matrix of this estimator is given in appendix~\ref{app:cov}. 

Finally in the high redshift range, $1\leq z\leq 1.7$, we split the 7.8 millions ELG's into two equal populations, with the same two free bias parameters $b_4$ and $b_5$. 

\begin{table}[t]
\centering
\begin{tabular}{ | c | c  | c | c | c | }
\hline
&$0\leq z\leq 0.5$ & $0.6\leq z\leq 1$ &$1\leq z\leq 1.7$ & combined \\ \hline
$\Gamma_0$ &$3.2$ & $2.4$ &$10.8$ &1.9 \\ \hline
$\mu_0$ &$3.2\times 10^{-3}$ & $2.7\times 10^{-3}$ &$5.2\times 10^{-3}$ & $1.8\times 10^{-3}$ \\ \hline
\end{tabular}
\caption{Constraints on $\Gamma_0$ and $\mu_0$, marginalised over the other parameters, obtained from the three redshift ranges in DESI.}
\label{tab:DESI}
\end{table}

In each case we calculate the cross-correlation dipole, as well as the monopole, quadrupole and hexadecapole of each population, in thin redshift bins of width $\Delta z=0.1$. We neglect correlations between different multipoles, but we take into account the correlation between the same multipole of different populations. We use the range of separation $10\leq d\leq 200\mph$. 
We assume that $\Theta=0$, so we have in total 7 free parameters: $\Gamma_0, \mu_0, b_1, b_2, b_3, b_4$ and $b_5$. The marginalised constraints on $\Gamma_0$ and $\mu_0$ are summarised in table~\ref{tab:DESI}. We see that the combined constraint on $\Gamma_0$ reaches $\Delta\Gamma_0=1.9$. This is 7 times larger than the constraint expected from the SKA, but it would provide the very first test of the equivalence principle at cosmological scales. Since we have never directly observed dark matter particles fall into a gravitational potential, a value of order unity for $\Gamma_0$ is not excluded.

%%%%%%%%%%%%%%%%%%%%%%%%%%%%
\section{Conclusion}
\label{sec:conclusion}

In this paper, we have shown that relativistic effects can be used to test the equivalence principle directly. We have used the dipole in the cross-correlation function of bright and faint galaxies to constrain deviations from Euler's equation. Such deviations would modify the relation between the galaxy velocity $V$ and the time component of the metric $\Psi$. They are therefore unconstrained by the monopole, quadrupole and hexadecapole of redshift-space distortions, since these observables are sensitive to the density and velocity, but not to the metric. Adding information from gravitational lensing is not enough, since lensing is affected by the sum of the two metric potentials, $\Phi+\Psi$, and not by $\Psi$ individually. The cross-correlation dipole provides therefore a unique opportunity to test for deviations in Euler's equation since it is \emph{directly} sensitive to $\Psi$ via the effect of gravitational redshift.

We have found that future surveys like DESI and the SKA can provide meaningful constraints on deviations from Euler's equation. Comparing with previous forecasts on modifications of gravity, we find that our constraints are weaker by 1-3 orders of magnitude with respect to constraints on the growth rate of structure and on the anisotropic stress. This is due to the fact that the signal-to-noise of the dipole is significantly lower than that of the RSD multipoles and of gravitational lensing, that are used to constrain those deviations. Hence the equivalence principle will always be more difficult to constrain than the growth of structure or the relation between the metric potentials. On the other hand, since we have never observed dark matter directly, any constraint obtained from the dipole would provide a significant improvement in our knowledge of this component.  

Finally, let us emphasise that the specificity of our work is to test the equivalence principle in a model-independent way. The parametrisation that we use is motivated by the example of scalar-tensor theories and Einstein-\ae ther theories, but it is in reality completely general. Indeed, at no point in our analysis we need to refer to a specific class of Lagrangian or theory. Our framework allows us to test for any deviation in Euler's equation, independently of the underlying theory. If on the other hand one chooses to start from a specific Lagrangian, with a set of free functions, then the rules of the game completely change. These free functions would indeed modify in a specific way the growth of structure, the relation between the metric components and Euler's equation. One can then \emph{indirectly} test for deviations in Euler's equation using RSD and lensing, through a measurement of these free functions, as for example in~\cite{Gleyzes:2015rua}. Such an approach would deliver more stringent constraints, since these free functions are very well measured via the growth rate, but it has the disadvantage that it is related to a specific class of model. 

Therefore, the real usefulness of relativistic effect is not to improve the constraints on parameters that can be measured via other methods. As pointed out in~\cite{Lorenz:2017iez}, relativistic effects are too small to have a measurable impact in this case. The usefulness of relativistic effects is rather to test for deviations that cannot be tested in a model-independent way by other methods. In this sense, relativistic effects are really complementary to standard observables and do add extra information.

%%%%%%%%%%%%%%%%%%%%%%%%%%%%

%%%%%%%%%%%%%%%%%%%%%%%%%%%%
\section*{Acknowledgements}
We thank Filippo Vernizzi, J\'{e}r\^{o}me Gleyzes, and Sergey Sibiryakov for clarifications about, respectively, scalar-tensor and Einstein-\ae ther theories. We also thank Ruth Durrer and Alex Hall for interesting discussions. We acknowledge support by the Swiss National Science Foundation.
%%%%%%%%%%%%%%%%%%%%%%%%%%%%

\appendix
%%%%%%%%%%%%%%%%%%%%%%%%%%%%
\section{The equations of Einstein-\ae ther gravity}
\label{appendix:Einstein-aether}
%%%%%%%%%%%%%%%%%%%%%%%%%%%%

In this appendix, we give all the equations of the Einstein-\ae ther gravity that were necessary to derive eq.~\eqref{eq:modified_Euler_Einstein-aether}. In most of this appendix, we keep full generality for the parameters~$c_{1\ldots 4}$ of the theory. Our equations can be compared with~\cite{2012JCAP...10..057B}, where a comprehensive analysis of the Einstein-\ae ther model with coupling to dark matter was performed.

\subsection{Equations of motion}

Recall that the action for gravity and \ae ther is
\begin{equation}
S = \frac{1}{16\pi G} \int  \dd^4 x \sqrt{-g} 
						\pac{ R + K\indices{^\mu^\nu_\rho_\sigma} \nabla_\mu u^\rho \nabla_\nu u^\sigma
								+ \lambda (u^\mu u_\mu + 1) }
\end{equation}
with 
$
K_{\mu\nu\rho\sigma} = c_1 g_{\mu\nu} g_{\rho\sigma}
										 + c_2 g_{\mu\rho} g_{\nu\sigma} 
										 + c_3 g_{\mu\sigma} g_{\nu\rho}
										 - c_4 u_{\mu} u_{\nu} g_{\rho\sigma},
$,
and $\lambda$ being a Lagrange multiplier, while the action of a dark point particle with bare mass $m$ is
\begin{align}
S_1
&= -m \int \dd \tau \; F(\gamma) \label{eq:action_pp_1}\\
&= -m \int \dd^4 x \int \dd\sigma \; \delta[x^\mu - x\e{p}^\mu(\sigma)]
							\sqrt{-g_{\mu\nu}\ddf{x\e{p}^\mu}{\sigma}\ddf{x\e{p}^\nu}{\sigma}} \,
							F\pa{\frac{-g_{\mu\nu} u^\mu\ddf{x^\nu\e{p}}{\sigma}}
											{ \sqrt{-g_{\mu\nu}\ddf{x\e{p}^\mu}{\sigma}\ddf{x\e{p}^\nu}{\sigma}} }
									}, \label{eq:action_pp_2}
\end{align}
where, in eq.~\eqref{eq:action_pp_1}, $\tau$ is the particle's proper time and $\gamma\define -u^\mu v_\mu$ is its Lorentz factor in the \ae ther frame; in eq.~\eqref{eq:action_pp_2}, $\sigma$ is an arbitrary parameter along the particle's worldline~$x\e{p}(\sigma)$. This latter form, albeit more complicated, is necessary to identify where $S_1$ actually depend on $g_{\mu\nu}$ and $u^\mu$.

\paragraph{Dark matter.} There are two ways of deriving the equation of motion for dark matter: either by directly differentiating $S_1$ with respect to $x^\mu\e{p}$, or from a Noether-like calculation, using the invariance of $S_1$ under diffeomorphisms. We choose here the second method as it will also bring energy conservation. Considering that $S\e{DM}$ is the superposition of many $S_1$, such that there are $n=\rho/m$ particles per unit volume, then the stress-energy tensor of dark matter is found to read
\begin{equation}\label{eq:stress-energy_DM_Einstein-aether}
T\e{DM}^{\mu\nu}
\define \frac{2}{\sqrt{-g}} \frac{\delta S\e{DM}}{\delta g_{\mu\nu}}
= (F-\gamma F_{,\gamma}) \rho v^\mu v^\nu + 2 \rho F_{,\gamma} u^{(\mu} v^{\nu)},
\end{equation}
with the symmetrisation convention $(\mu\nu)\define (\mu\nu + \nu\mu)/2$. In the functional derivative of~\eqref{eq:stress-energy_DM_Einstein-aether} we assumed that $u^\mu$ is the fundamental \ae ther variable.\footnote{One could make another choice and consider that $u_\mu$ is the fundamental variable. In this case the result would not contain $2 \rho F_{,\gamma} u^{(\mu} v^{\nu)}$. This term would then have to come from the \ae ther action. The important thing is to choose a convention and follow it consistently.} The equation of motion is then
\begin{equation}
\nabla_\mu (T\e{DM})\indices{^\mu_\nu}
= \rho F_{,\gamma} v^\mu \nabla_\nu u_\mu + \nabla_\mu (\rho F_{,\gamma} v_\nu u^\mu)
\end{equation}
which yields the system given in sec.~\ref{sec:Einstein-aether},
\begin{align}
\nabla_\mu (\rho v^\mu) &= 0 \\
v^\nu \nabla_\nu \pac{(F-\gamma F_{,\gamma})v^\mu}
&= F_{,\gamma} \omega\indices{^\mu_\nu} v^\nu - \dot{\gamma} F_{,\gamma\gamma} u^\mu ,
\end{align}
with $\omega_{\mu\nu}\define \nabla_\mu u_\nu - \nabla_\nu u_\mu$, and a dot is a derivative with respect to proper time.

\paragraph{Aether.} The equation of motion for \ae ther is obtained by directly differentiating the total action with respect to $u^\mu$, which yields
\begin{equation}
\nabla_\nu (K\indices{^\sigma^\nu_\rho_\mu}\nabla_\sigma u^\rho) 
= -c_4 u^\nu \nabla_\nu u^\rho \nabla_\mu u_\rho
	+ \lambda u_\mu + 8\pi G \rho v_\mu F_{,\gamma}.
\end{equation}
Defining $J\indices{^\mu_\rho} \define K\indices{^\mu^\nu_\rho_\sigma} \nabla_\nu u^\sigma$, and using the constraint $u^\mu u_\mu = -1$, we can eliminate the Lagrange multiplier and get
\begin{equation}\label{eq:EOM_aether}
h_{\mu\rho} \nabla_\nu J^{\nu\rho} = -c_4 h^\sigma_\mu (u^\nu \nabla_\nu u^\rho \nabla_\sigma u_\rho) 
																+ 8\pi G\rho F_{,\gamma} \, h_{\mu\nu} v^\nu,
\end{equation}
where $h_{\mu\nu}\define g_{\mu\nu}+u_\mu u_\nu$ is the spatial metric in the rest frame of \ae ther.

As for its stress-energy tensor, a long but straightforward calculation, still assuming that $u^\mu$ is the independent variable, yields
\begin{align}\label{eq:stress-energy_aether}
8\pi G T_{\mu\nu}\h{\ae}
&\define -\frac{1}{\sqrt{-g}}\frac{\delta}{\delta g^{\mu\nu}} \int \dd^4 x \, \sqrt{-g} 
																		\pac{ K\indices{^\mu^\nu_\rho_\sigma} \nabla_\mu u^\rho \nabla_\nu u^\sigma
																					+ \lambda (u^\mu u_\mu + 1)
																				}\\
&= c_1 \pa{\nabla_\rho u_\mu \nabla^\rho u_\nu - \nabla_\mu u^\rho \nabla_\nu u_\rho} 
	- c_4 u^\rho u^\sigma \nabla_\rho u_\mu \nabla_\sigma u_\nu \\
	&\quad- \nabla_\rho \pac{ J_{(\mu\nu)} u^\rho + J\indices{^\rho_(_\mu} u_{\nu)} - u_{(\mu} J\indices{_\nu_)^\rho} }
	+ \lambda u_\mu u_\nu 
	+ \frac{1}{2} J\indices{^\rho_\sigma} \nabla_\rho u^\sigma g_{\mu\nu}.\\
&= c_1 \pa{\nabla_\rho u_\mu \nabla^\rho u_\nu - \nabla_\mu u^\rho \nabla_\nu u_\rho} 
		- c_4 u^\rho u^\sigma \pa{ \nabla_\rho u_\mu \nabla_\sigma u_\nu
														+ u_\mu u_\nu \nabla_\rho u^\lambda \nabla_\sigma u_\lambda } \\
		&\quad - \nabla_\rho \pac{ J_{(\mu\nu)} u^\rho + J\indices{^\rho_(_\mu} u_{\nu)} - u_{(\mu} J\indices{_\nu_)^\rho} }
		- u^\rho \nabla_\sigma J\indices{^\sigma_\rho} u_\mu u_\nu 
		+ \frac{1}{2} J\indices{^\rho_\sigma} \nabla_\rho u^\sigma g_{\mu\nu} \\
		&\quad - 8\pi G\gamma\rho F_{,\gamma} u_\mu u_\nu
\end{align}

\paragraph{Gravity.} The Einstein field equation is, as usual, obtained by differentiating the total action (with many dark matter particles) with respect to the metric, and simply reads
\begin{equation}
R_{\mu\nu} - \frac{1}{2} R g_{\mu\nu} = 8\pi G (T_{\mu\nu}\h{\ae}+T_{\mu\nu}\h{DM}),
\end{equation}
in which one just has to plug the expressions of the stress-energy tensors obtained in the previous paragraphs. The result for the total stress-energy tensor is
\begin{align}
8\pi G (T_{\mu\nu}\h{\ae}+T_{\mu\nu}\h{DM})
&= c_1 \pa{\nabla_\rho u_\mu \nabla^\rho u_\nu - \nabla_\mu u^\rho \nabla_\nu u_\rho} 
		- c_4 \pa{ a^\rho a_\rho u_\mu u_\nu + a_\mu a_\nu } \nonumber\\
		&\quad - \nabla_\rho \pac{ J_{(\mu\nu)} u^\rho + J\indices{^\rho_(_\mu} u_{\nu)} - u_{(\mu} J\indices{_\nu_)^\rho} }
		- u^\rho \nabla_\sigma J\indices{^\sigma_\rho} u_\mu u_\nu 
		+ \frac{1}{2} J\indices{^\rho_\sigma} \nabla_\rho u^\sigma g_{\mu\nu} \nonumber\\
		&\quad + 8\pi G\rho \pac{(F-\gamma F') v_\mu v_\nu +F_{,\gamma} \pa{2u_{(\mu}v_{\nu)}-u_\mu u_\nu} },
\end{align}
with $a^\mu \define u^\nu\nabla_\nu u^\mu$. Note that our total stress-energy tensor looks different from eq.~(82) of \cite{2012JCAP...10..057B}. This is due to our different conventions for the signature of the metric and for the sign of $c_4$. One can check that, using the equation of motion of \ae ther, and re-introducing the Lagrange multiplier, that our results agree.

\subsection{Cosmology}

\subsubsection{Background}

First consider an FLRW space-time. The symmetries of this solution impose that the \ae ther flow reads~$u^\mu=\delta^\mu_t$, where $t$ denotes cosmic time. The dynamics of dark matter and \ae ther are thus trivial, and the only degree of freedom is the scale factor. Plugging the solution into Einstein's equation, one finds the modified Friedmann equations
\begin{align}
\pa{1-\frac{c_1+3c_2+c_3}{2}} H^2
&= \frac{8\pi G\rho}{3} - \frac{\mathcal{K}}{a^2} \\
\pa{1-\frac{c_1+3c_2+c_3}{2}} \frac{\ddot{a}}{a}
&= -\frac{4\pi G\rho}{3},
\end{align}
where $\mathcal{K}$ is the spatial curvature of the homogeneity hypersurfaces. We could have added a cosmological constant to the right-hand side by adding the corresponding term in the action.

The apparition of the factor $1-(c_1+3c_2+c_3)/2$ in the ``kinetic side'' of the Friedmann equations can be understood as follows. Geometrically, $\ddot{a}/a, H$ are related to the extrinsic curvature~$K_{\mu\nu}=\nabla_\mu u_\nu=H h_{\mu\nu}$ of the homogeneity hypersurfaces, with $h_{\mu\nu}=g_{\mu\nu}+u_\mu u_\nu$, $u^\mu$ being the unit normal to the hypersurfaces. Now using the $3+1$ decomposition of the Ricci scalar, we can rewrite the Einstein-\ae ther Lagrangian as
\begin{align}
R + K\indices{^\mu^\nu_\rho_\sigma} \nabla_\mu u^\rho \nabla_\nu u^\sigma
&= {}^3R + K^{\mu\nu} K_{\mu\nu} - (K^\mu_\mu)^2 + (c_1+c_3) K^{\mu\nu} K_{\mu\nu} + c_2 (K^\mu_\mu)^2 \\
&= {}^3R - 6 H^2 \pa{ 1 - \frac{c_1+3c_2+c_3}{2} } \\
&= {}^3R + \pa{ 1 - \frac{c_1+3c_2+c_3}{2} } \pac{K^{\mu\nu} K_{\mu\nu} - (K^\mu_\mu)^2}
\end{align}
Hence the intrinsic-curvature part is unchanged, but the extrinsic-curvature part is renormalised, whence the result in the Friedmann equations.

\subsubsection{Perturbations}

We now consider scalar cosmological perturbations about the FLRW background in the Poisson gauge. The metric is taken to be the same as in eq.~\eqref{eq:metric}, and the four-velocity of \ae ther and dark matter are given, respectively, by $u^\mu=\bar{u}^\mu+\delta u^\mu$ and $v^\mu=\bar{u}^\mu+\delta v^\mu$. Their normalisation implies~$\delta u_0=\delta v_0 = -a\Phi$, and we introduce the velocity potentials such that~$\delta u_i=a\partial_i U$ and $\delta v_i = a\partial_i V$. As usual, the dark matter density perturbation is parameterised as~$\rho = \bar{\rho}(1+\delta)$.

\paragraph{Dark matter.} While the continuity equation for dark matter remains unchanged,
\begin{equation}
\delta' + \Delta V - 3\Phi' = 0,
\end{equation}
with $\Delta = \delta^{ij}\partial_i\partial_j$, the modified Euler equation~\eqref{eq:EOM_DM_Einstein-aether} reads
\begin{equation}
(1-Y) (V'+\Hc V + \Psi) = -Y\pa{U'+\Hc U + \Psi},
\end{equation}
or $A_V = Y(A_V-A_U)$ if we introduce the four-acceleration potential~$A_X\define X'+\Hc X + \Psi$, and the coupling constant~$Y\define F_{,\gamma}(1)$, as stated in the main text.

\paragraph{Aether.} At linear order, the equation of motion~\eqref{eq:EOM_aether} is shown to read
\begin{multline}\label{eq:EOM_aether_perturbed}
-(c_1+c_4) \pa{ A_U' + \mathcal{H} A_U } + (c_1+c_2+c_3) \Delta U \\
+ (c_1+3c_2+c_3) \pac{ \pa{\frac{a''}{a}-2\mathcal{H}^2} U - \Phi' - \mathcal{H}\Psi }
= 8\pi G a^2\bar{\rho} Y (V-U).
\end{multline}
When we take into account the observational constraints $c_1+c_3=c_1+c_4=0$, it reduces to
\begin{equation}\label{eq:EOM_aether_perturbed_2}
c_2\Delta U
- 3 c_2 \pac{ \pa{2\mathcal{H}^2-\frac{a''}{a}} U + \Phi' + \mathcal{H}\Psi }
= 8\pi G a^2\bar{\rho} Y (V-U).
\end{equation}

\paragraph{Gravity.} Finally, the components ${}^0_0$, ${}^0_i$, and the trace-less part of ${}^i_j$ of the Einstein field equation yield respectively
\begin{gather}
\Delta \Phi - 3\Hc(\Phi'+\Hc\Psi) 
= 4\pi G a^2\bar{\rho} \delta
	- \frac{1}{2}(c_1+c_4) \Delta A_U + \frac{1}{2}(c_1+3c_2+c_3)\Hc\pac{ \Delta U - 3(\Phi'+\Hc\Psi) } \\
\Phi'+\Hc\Psi 
= -4\pi G a^2\bar{\rho} V - \frac{1}{2} (c_1+c_4) (A_U'+\Hc A_U)
- \frac{1}{2}(c_1+3c_2+c_3)\pa{2\Hc^2-\frac{a''}{a}} U \\
\pa{\partial_i\partial_j - \frac{1}{3}\Delta}(\Psi-\Phi)
= (c_1+c_3) \pa{\partial_i\partial_j - \frac{1}{3}\Delta} (U'+\Hc U),
\end{gather}
which, for $c_1+c_3=c_1+c_4=0$, becomes
\begin{align}
\Delta \Phi - 3\Hc(\Phi'+\Hc\Psi) &= 4\pi G a^2 \bar{\rho}\delta + \frac{c_2}{2}\Hc\pac{ \Delta U - 3(\Phi'+\Hc\Psi) }
\label{eq:EFE_AE_pert_1}\\
\Phi'+\Hc\Psi &= -4\pi G a^2\bar{\rho} V - \frac{c_2}{2}\pa{2\Hc^2-\frac{a''}{a}}U 
\label{eq:EFE_AE_pert_2} \\
\Psi &= \Phi
\label{eq:EFE_AE_pert_3}
\end{align}

\paragraph{Modified Euler equation.} Let us use the whole set of the above equations to simplify the modified Euler equation. In particular, we would like to rewrite the relative acceleration potential $A_V-A_U$ in terms of $V$ only. For that purpose, we use the background dynamics, which implies in particular
\begin{equation}
\frac{a''}{a} - 2\Hc^2 = -\frac{8\pi G \bar{\rho} a^2}{2-3c_2},
\end{equation}
which we inject into eq.~\eqref{eq:EFE_AE_pert_2}, and then inject the latter into the equation of motion~\eqref{eq:EOM_aether_perturbed_2},
\begin{equation}
c_2 \Delta U - 12\pi c_2 G a^2 \hat{\rho} \pa{ \frac{2-c_2}{2-3 c_2} U - V } = 8\pi Y G a^2\hat{\rho} (V-U).
\end{equation}
In Fourier space, this can be rewritten as $V-U = h(a,k) V$, where
\begin{equation}
h(a,k) \define \frac{c_2(2-3 c_2) a k^2 + 9 c_2^2 \Omega\e{m0} H_0^2}
								{c_2 (2-3 c_2) a k^2 - \Omega\e{m0} H_0^2 \pac{(2-3 c_2)Y+3c_2(1-c_2/2)}},
\end{equation}
from which we deduce
\begin{equation}
A_V - A_U = h(V'+\Hc V)+h' V,
\end{equation}
and the modified Euler equation finally reads
\begin{equation}
V'+ \frac{1-Y h - \Hc^{-1}Y h'}{1-Y h} \, \Hc V + \frac{\Phi}{1-Y h} = 0.
\end{equation}
For $k\ll H_0$, $h(a,k)\approx 1$, and the above reduces to
\begin{equation}
V'+ \Hc V + \frac{\Phi}{1-Y} = 0 ,
\end{equation}
which has the same for as eq.~\eqref{eq:parametrisation_modified_Euler}, with $\Theta=0$ and $\Gamma=Y/(1-Y)$.

%%%%%%%%%%%%%%%%%%%%%%%%%%%%%%
\section{Redshift corrections and equivalence principle}
\label{app:equivalence}
%%%%%%%%%%%%%%%%%%%%%%%%%%%%%%

\begin{figure}[h!]
\centering
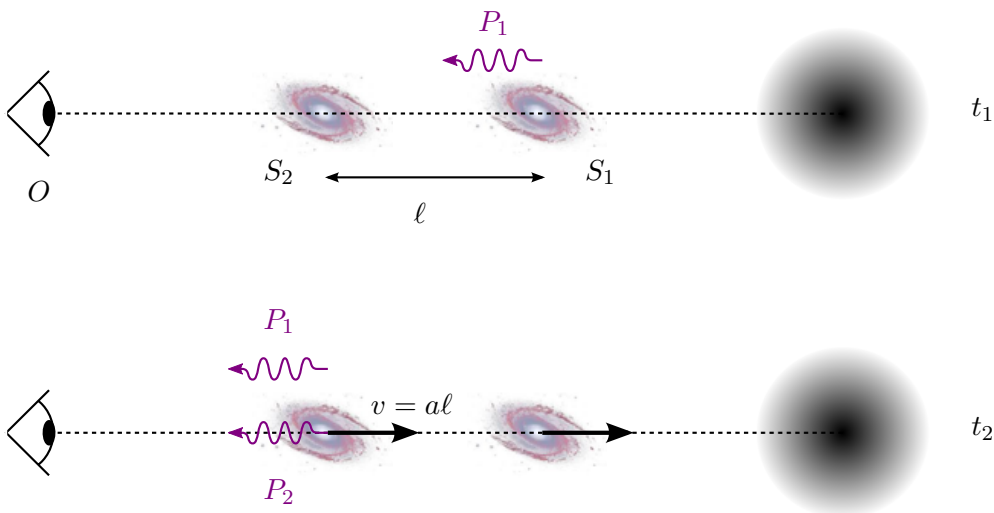
\caption{Two light sources~$S_1$, $S_2$, close to a massive object, are at rest and separated by a distance~$\ell$ at $t=t_1$. A photon $P_1$ is emitted towards the observer~$O$ by $S_1$ at that instant, which passes $S_2$ at $t_2\approx t_1+\ell$, when the latter emits, in turn, a photon~$P_2$.}
\label{fig:equivalence}
\end{figure}

This appendix aims at emphasising the fundamental link between the cancellation of some relativistic corrections to the galaxy number counts~$\Delta$ and the equivalence principle. For that purpose, let us forget about cosmology, and consider the simple case of two light sources in radial free fall onto a massive object, as depicted in fig.~\ref{fig:equivalence}, and discuss how an observer, at rest far from them, sees the light they emit.

We first work in the observer's frame. For simplicity, suppose that both sources are initially at rest, separated by a distance~$\ell$. At $t=t_1$, the source $S_1$ closest to the massive object, undergoing a gravitational potential~$\Psi_1$ emits a photon $P_1$ with frequency~$\omega\e{s}$ in its own frame. This photon reaches the source $S_2$ at time~$t_2\approx t_1+ \ell$, where we have neglected the velocity of $S_2$ with respect to the speed of light. This velocity, however, is not zero, because $S_2$ falls in the gravitational field of the massive object, with an acceleration $a$, and hence during $P_1$'s propagation time, it gained a velocity~$v=a\ell$.

Now suppose that, at $t_2$, $S_2$ emits a photon $P_2$ with frequency~$\omega\e{s}$. Then, in principle, this frequency differs from the frequency~$\omega_1$ of $P_1$ in the frame of $S_2$, because of both Doppler and Einstein effects. Namely, at lowest order,
\begin{equation}\label{eq:redshift_correction_S1_S2}
\delta z = \frac{\omega_1-\omega\e{s}}{\omega\e{s}} = v + \Psi_1-\Psi_2 = \pa{a+\partial_r \Psi} \ell.
\end{equation}
Therefore, everything happens as if the source $S_2$ were sending simultaneously two photons with different frequencies. Whatever happens to those photons, their mutual redshift~$\delta z$ is preserved until they reach $O$. In an expanding Universe, this reasoning still applies if one replaces~$a$ by the peculiar acceleration. Once written in comoving coordinates, the corresponding redshift correction reads
\begin{equation}
\delta z = ( \vect{V}'\cdot\vect{n} + \Hc \vect{V}\cdot\vect{n} + \partial_r \Psi) \ell ,
\end{equation}
where we recognise three terms of the six terms of $\Delta\h{rel}$, as written in eq.~\eqref{eq:Delta_relativistic}. 

Of course, since the sources are in radial free fall, $a=-\partial_r\Psi$, and the two effects contributing to eq.~\eqref{eq:redshift_correction_S1_S2} cancel exactly. This cancellation must be understood as a consequence of Einstein's equivalence principle. Indeed, let us consider the above situation in the rest frame of $S_2$. In that frame, the effects of gravity are eliminated, modulo curvature terms, so $S_1$ appears to be at rest, while light propagates as in Minkowski space-time. Consequently, the frequency of $P_1$ as observed by $S_2$ is unchanged with respect to its emission frequency. The only way to escape this reasoning is to violate the equivalence principle, either by assuming that (i) $S_1$ and $S_2$ fall differently, or (ii) $P_1$ falls differently from $S_1$ and $S_2$. In this article, we considered possible departures from (ii), where dark matter and photons are not identically coupled to gravitation.

%%%%%%%%%%%%%%%%%%%%%%%%%%%%%%
\section{Covariance matrices}
\label{app:cov}
%%%%%%%%%%%%%%%%%%%%%%%%%%%%%%

We need 7 covariance matrices: the covariance of the cross-correlation dipole $C^1_{\B\F}$, the covariance of the monopole $C^0_{\rm X}$ (where $\rm{X}=\B,\F$), quadrupole $C^2_{\rm X}$ and hexadecapole $C^4_{\rm X}$, and the cross-covariance of the bright and faint monopole $C^0_{\B\F}$, quadrupole $C^2_{\B\F}$ and hexadecapole $C^4_{\B\F}$. Note that $C^0_{\B\F}$ does not denote the covariance of the cross-correlation monopole between bright and faint galaxies, but rather the covariance between the monopole of the bright and the monopole of the faint. The covariance matrix for the monopole is then given by
\be
C^0=\left(\begin{array}{cc} C^0_\B&C^0_{\B\F}\\C^0_{\B\F}&C^0_{\F} \end{array}\right)\, ,
\ee
and similarly for the quadrupole and the hexadecapole.

The covariance matrices contain three contributions: a Poisson contribution due to the fact that we observe a finite number of galaxies, a cosmic variance contribution due to the fact that we observe a finite volume and a mixed contribution from Poisson and cosmic variance. Since the standard terms (density and RSD) are always larger than the relativistic terms, we can neglect the contribution from the relativistic terms to the cosmic variance and to the mixed contribution. The only exception is for the covariance of the dipole, where the cosmic variance contribution from the standard terms vanishes. However, the mixed contribution from the standard terms does not vanish and always dominates over the relativistic cosmic variance contribution (see fig. 2 of~\cite{Hall:2016bmm}), so we can safely neglect the latter also in this case. We follow the method developed in~\cite{Bonvin:2015kuc,Hall:2016bmm} to calculate the covariance matrices. We obtain
\begin{align}\mathcal{V}
C^1_{\B\F}=&\frac{3}{4\pi \bar n_\B \bar n_\F \bar N^2 \mathcal{V} \ell_{\rm p}  d_i^2}\delta_{ij}\nonumber\\
&+\frac{1}{\bar N \mathcal{V}}\left[\frac{1}{4\bar n_\F}\left(\frac{b_\B^2}{3}+\frac{2b_\B f}{5}+\frac{f^2}{7} \right)
+\frac{1}{4\bar n_\B}\left(\frac{b_\F^2}{3}+\frac{2b_\F f}{5}+\frac{f^2}{7} \right) \right]G_{\ell=1}(d_i,d_j)\, , \label{cov1}\\
C^0_\X=&\frac{1}{2\pi \bar N^2 \mathcal{V} \ell_{\rm p}  d_i^2}\delta_{ij}+\frac{1}{\mathcal{V}}\left(b_\X^4+\frac{4b_\X^3f}{3}+\frac{6b_\X^2f^2}{5}+\frac{4b_\X f^3}{7}+\frac{f^4}{9} \right) D_{\ell=0}(d_i,d_j)\nonumber\\
&+\frac{1}{\bar N \mathcal{V}}\left(b_\X^2+\frac{2b_\X f}{3}+\frac{f^2}{5} \right) G_{\ell=0}(d_i,d_j)\, ,\label{cov0}\\
C^2_\X=&\frac{5}{2\pi \bar N^2 \mathcal{V} \ell_{\rm p}  d_i^2}\delta_{ij}+\frac{1}{\mathcal{V}}\left(\frac{44b_\X^3f}{105}+\frac{18b_\X^2f^2}{35}+\frac{68b_\X f^3}{231}+
\frac{83f^4}{1287} \right) D_{\ell=2}(d_i,d_j)\nonumber\\
&+\frac{1}{\bar N \mathcal{V}}\frac{1}{5}\left(b_\X^2+\frac{22b_\X f}{21}+\frac{3f^2}{7} \right) G_{\ell=2}(d_i,d_j)\, ,\label{cov2}\\
C^4_\X=&\frac{9}{2\pi \bar N^2 \mathcal{V} \ell_{\rm p} d_i^2}\delta_{ij}+\frac{1}{\mathcal{V}}\left(\frac{b_\X^4}{9}+\frac{52b_\X^3f}{231}+\frac{1286b_\X^2f^2}{5005}+\frac{436b_\X f^3}{3003}+\frac{79f^4}{2431} \right) D_{\ell=4}(d_i,d_j)\nonumber\\
&+\frac{1}{\bar N \mathcal{V}}\frac{1}{3}\left(\frac{b_\X^2}{3}+\frac{26b_\X f}{77}+\frac{643f^2}{5005} \right) G_{\ell=4}(d_i,d_j)\, ,\label{cov4}\\
C^0_{\B\F}=&\frac{1}{\mathcal{V}}\left[b_\B^2b_\F^2+\big(b_\B^2 b_\F +b_\F^2 b_\B\big)\frac{2f}{3} +\big(b_\B^2+4b_\B b_\F +b_\F^2 \big)\frac{f^2}{5}
+\big(b_\B+b_\F \big)\frac{2f^3}{7}+\frac{f^4}{9} \right]\nonumber\\
&\quad\times D_{\ell=0}(d_i,d_j)\, ,\\
C^2_{\B\F}=&\frac{1}{\mathcal{V}}\left[\frac{b_\B^2b_\F^2}{5}+\big(b_\B^2 b_\F +b_\F^2 b_\B\big)\frac{22f}{105} +\big(b_\B^2+4b_\B b_\F +b_\F^2 \big)\frac{3f^2}{35}+\big(b_\B+b_\F \big)\frac{34f^3}{231}+\frac{83f^4}{1287} \right]\nonumber\\
&\quad\times D_{\ell=2}(d_i,d_j)\, ,\\
C^4_{\B\F}=&\frac{1}{\mathcal{V}}\left[\frac{b_\B^2b_\F^2}{9}+\big(b_\B^2 b_\F +b_\F^2 b_\B\big)\frac{26f}{231} +\big(b_\B^2+4b_\B b_\F +b_\F^2 \big)\frac{643f^2}{15015}+\big(b_\B+b_\F \big)\frac{218f^3}{3003}+\frac{79f^4}{2431} \right]\nonumber\\
&\quad\times D_{\ell=2}(d_i,d_j)\, .
\end{align}
Here $\mathcal{V}$ denotes the volume of the survey or of the redshift bin of interest, $\bar N$ is the number density, $\bar n_\B$ and $\bar n_\F$ are the fraction of bright and faint galaxies, and $\ell_{\rm p}$ is the size of the cubic pixel in which $\Delta$ is measured. We use $\ell_{\rm p}=2$\,Mpc$/h$. The functions $D_\ell$ and $G_\ell$ are defined as follows for $\ell=0,1,2,4$
\begin{align}
D_\ell(d_i,d_j)=&\frac{(2\ell+1)^2}{\pi^2}\int \dd k \; k^2 P_{\delta\delta}^2(\bar z, k)j_\ell(kd_i)j_\ell(kd_j)\, ,\\
G_\ell(d_i,d_j)=&\frac{2(2\ell+1)^2}{\pi^2}\int \dd k \; k^2 P_{\delta\delta}(\bar z, k)j_\ell(kd_i)j_\ell(kd_j)\, .
\end{align}

In the forecasts for DESI, we correlate three populations of galaxies A, B and C in the middle redshift range. We optimise the signal-to-noise in this range by weighting each pair of pixels by the bias difference between the populations, as discussed in~\cite{Bonvin:2015kuc}. The covariance of this new estimator is then given by
\begin{align}
C^1_{\rm{ABC}}=&\frac{3}{4\pi\bar N^2 \mathcal{V} \ell_{\rm p}  d_i^2}\sum_{L,L'={\rm A,B,C}}\frac{(b_L-b_{L'})^2}{\bar n_L\bar n_{L'}}\cdot \delta_{ij} \label{cov13pop}\\
&+\frac{1}{4\bar N \mathcal{V}}\sum_{\substack{L,L',L''\\={ \rm A, B, C}}}\frac{1}{\bar n_{L'}}(b_L-b_{L'})(b_{L''}-b_{L'})
\left[\frac{b_L b_{L''}}{3}+\big(b_L+b_{L''} \big)\frac{f}{5}+\frac{f^2}{7} \right]G_{\ell=1}(d_i,d_j)\, .\nonumber
\end{align}

\bibliographystyle{JHEP.bst}
\bibliography{bibliography_Euler.bib}

\end{document}